\documentclass[12pt,reqno, times]{article}
\usepackage[english]{babel}
\usepackage[backend=biber, natbib, maxcitenames=5, mincitenames=1, style=apa]{biblatex}
\usepackage{csquotes}
\usepackage{enumerate}
\usepackage{setspace}
\usepackage[margin=1.00in]{geometry}
\usepackage{authblk}
\usepackage{graphicx}
\usepackage{palatino}
\usepackage{xcolor}
\usepackage{amsfonts, amsmath, latexsym, amssymb}
\usepackage{mathabx}
\usepackage{amsthm}
\usepackage[titletoc]{appendix} 
\newtheorem{prop}{Proposition}
\newtheorem{corollary}{Corollary}

\newtheorem{lemma}{Lemma}

\newcommand{\R}{\mathbb{R}}
\newcommand{\N}{\mathcal{N}}

\newcommand{\enum}{\begin{enumerate}}
\newcommand{\fenum}{\end{enumerate}}
\newcommand{\enlist}{\begin{itemize}}
\newcommand{\fenlist}{\end{itemize}}
\newcommand{\equa}{\begin{eqnarray}}
\newcommand{\fequa}{\end{eqnarray}}
\newcommand{\equann}{\begin{eqnarray*}}
\newcommand{\fequann}{\end{eqnarray*}}
\usepackage{float}
\usepackage{comment} 
\includecomment{comment}
\excludecomment{comment} 
\usepackage{enumitem}

\newcommand{\E}{\mathbb{E}}
\newcommand{\Var}{\operatorname{Var}}

\title{Confirmation Bias as Rational Mistrust}
\author{Gabriel Martinez\thanks{Unaffiliated for research purposes.} ~~  Nicholas H. Tenev\thanks{Office of the Comptroller of the Currency, USA. The views expressed in this paper are my own and do not represent the views of the OCC, the Department of the Treasury, or the United States government. This paper is the result of my independent research and has not been reviewed by the OCC. Corresponding author: nicholas.tenev@protonmail.ch; ORCID 0000-0001-9403-8664. We are grateful to Nathan Yoder, Marzena Rostek, and seminar participants at the 2023 Econometric Society North American Summer Meeting and 2020 Delhi School of Economics Winter School for their insightful comments. Declarations of interest: none.} 
}
\date{September 25, 2026\footnote{First version October 3, 2020. Previously titled ``Optimal Echo Chambers.''}}
\begin{document}

\maketitle

\begin{abstract}
	When learning from others, people tend to focus their attention on those with similar views. This is often attributed to flawed reasoning, and thought to slow learning and polarize beliefs. However, we show that this confirmatory bias is a rational response to uncertainty about the accuracy of information sources, and can improve learning and reduce disagreement among those with diverse beliefs. Furthermore, extending the range of views someone is exposed to can backfire, slowing their learning by reducing trust. We model a Bayesian decision maker who chooses an acceptable range of content, and then observes one report drawn from those that fall within that range. With uncertainty about which sources are accurate, excluding opposing views can be beneficial, as they are more likely to come from inaccurate sources. The optimal range of views to pay attention to balances the credibility of views similar to one's own against the usefulness of those further away.
\end{abstract}


\thispagestyle{empty}
\newpage 
\setcounter{page}{1}
\section{Introduction}

Learning from others is fundamental to making decisions. Yet when deciding whom to learn from, people often choose those with similar beliefs. Consumers of news from both traditional and social media prefer sources aligned with their politics, and even when money is on the line, investors tend to follow social media accounts with similar views on the market.\footnote{\citet{iyengar2009red} show that people prefer news sources aligned with their politics. \citet{bakshy2015exposure} find that Facebook users' friends tend to share their politics, and thus tend to see news they agree with. \citet{garimella2018political} find similar selective exposure on Twitter; see also \citet{gorodnichenko2021social}. \citet{cookson2020echo} find that investors on social media tend to follow others whose sentiment (bullish/bearish) matches their own. \citet{stroud2017selective} reviews communications literature on selective exposure.} This form of confirmation bias---people listening only to those they agree with---is often attributed to cognitive bias, at odds with accuracy. It has been blamed for polarization in beliefs and the spread of misinformation during a global pandemic, and framed as a threat to democracy as well as the stability of financial markets.\footnote{See (respectively) \citet{faia2021biases} (discussed further in footnote \ref{fn:faia}), \citet{chou2018addressing}, \citet{sunstein2018republic}, and \citet{tang2017echo}.} 

We provide an alternative explanation for confirmation bias that is fully rational and matches the empirical evidence well, yet has very different policy implications than cognitive bias. If some information sources are less accurate than others but it is not clear which, it is reasonable to place more trust in those one agrees with. \citet{gentzkow2006media} show this is true for a binary state, but otherwise this insight remains largely absent from the literature, despite its relevance to a wide variety of learning environments.\footnote{\citet{levy2019echo} review the economics literature on echo chambers, but focus on segregation that does not improve the accuracy of beliefs. \citet{hart2009feeling} survey psychology work on selective exposure, weighing confirmation of pre-existing attitudes against accuracy as competing motives.} We extend this result to a richer context (a continuous state space), which reveals a novel tradeoff between the credibility of views close to one's own and the usefulness of those further away. Studying the optimal selection of information based on content in this environment generates counter-intuitive new insights that challenge the typically negative welfare implications of prior work. Our model can explain confirmation bias in both choosing which reports to pay attention to and deciding how much to believe them, and we apply our results to learning from news organizations as well as dialogue with peers.

Proposals to mitigate confirmation bias (and messaging campaigns more generally) often aim to expose people to views that challenge their beliefs---for example, appealing to people to diversify their sources themselves, or even giving them less control over what appears in their social media feeds.\footnote{While some suggestion algorithms show users news stories similar to those they already like, others such as the BuzzFeed News feature ``Outside Your Bubble'' \citep{smith2017} are designed to show users a range of perspectives. See \citet{obama2017} for an appeal to self-diversify.} It may seem natural that if somebody is misinformed and only paying attention to similar views, exposing them to more accurate information will bring their beliefs closer to the truth. But this doesn't always work in practice---for example, exposing partisans to opposing political views can result in wider disagreement, rather than bringing views together.\footnote{\citet{bail2018exposure} find that social media users incentivized to follow accounts they are politically opposed to polarize more than the control group. See Table \ref{tab:bail2018} in Appendix \ref{app:proofs} for further detail.} Our theory can help resolve this seeming paradox. We show that if someone is unsure which sources are reliable, broadening the views they are exposed to can backfire, undermining their trust in others and slowing their learning. 

In our model, an agent starts with a prior belief about the state of the world, which is real-valued. There exists a collection of signal sources, and some are more accurate and informative than others, but the agent is not sure which. The agent chooses a range of reports to sample from, and draws one from the subset of reports within that range of her prior expectation. Then, having observed the report (but not the quality of its source), she takes an action and is rewarded based on how closely it matches the true state.

The agent's ability to focus attention on a specific range of views can be thought of as choosing an echo chamber, and is key to our results.\footnote{We define echo chambers broadly, as environments in which people tend to encounter views similar to their own.} It applies to a wide variety of contexts including curating one's social media, picking a news source with a known partisan slant, or even deciding where to live (all discussed in Section \ref{sec:model}). Filtering out signals that challenge one's beliefs will sometimes require an editorial intermediary, though our results hold even if the agent's control over what she hears is inexact (Corollary \ref{cor:leaky}).\footnote{Another interpretation of such a ``leaky'' echo chamber is that the agent divides their attention across news sources, allocating more attention to some than others.} And even when no intermediary filter is available, we show that people have incentives to coordinate and learn from those with similar views (Corollary \ref{cor:netform}).

Our main results are as follows. If there exist information sources of low enough accuracy, reports close to one's prior expectation are more likely to be from accurate types (Proposition \ref{prop:hwithinr}). Accordingly, it can be optimal to restrict attention to viewpoints similar to one's own (Proposition \ref{prop:finiter}), as this screens out signals that are less accurate in expectation. This is true even if there are very few low-quality sources, and need not unravel if one knows one's sources are learning from echo chambers of their own (Section \ref{sec:secondary}). The underlying tension is a novel tradeoff between the credibility of confirmatory reports and the usefulness of those that challenge one's prior, which can be articulated cleanly given certain regularity assumptions (Corollary \ref{cor:creduseful}). We also find that reports close to the prior expectation can in some cases move beliefs further than more contrary views, since the former are more credible (Proposition \ref{prop:actionnonmon}). The message most persuasive to someone may not be the truth, but instead something closer to their own beliefs. Focusing on a limited range of viewpoints may thus do more to correct mistaken beliefs than exposure to opposing views, since only trusted opinions are heeded. Furthermore, confirmatory sampling moves the beliefs of a population closer to the truth on average (Corollary \ref{cor:polarization}), which can reduce disagreement in a population with sufficiently diverse beliefs. To be clear, ignoring opposing views can prevent those with extreme beliefs from learning the information most useful to them, and can increase the potency of misinformation (Corollary \ref{cor:ineq}). However, this may be impossible to remedy without first addressing the underlying friction: uncertainty about accuracy.

We further contextualize our contribution in Section \ref{sec:litreview}, before detailing the baseline model in Section \ref{sec:model}. Section \ref{sec:results} derives the main results, which require little in terms of regularity assumptions, and Section \ref{sec:normaldist} derives additional cleaner results by assuming normality. Section \ref{sec:pop} applies our results to the dispersion of beliefs and the network of attention in a population of agents. In Section \ref{sec:secondary}, we study secondary learning and argue that confirmation bias can be optimal even when one's sources are learning from echo chambers themselves. Section \ref{sec:conclusion} concludes by discussing policy implications and ideas for future work. Throughout, longer proofs and ancillary results are postponed to Appendix \ref{app:proofs}.

\section{Literature review}\label{sec:litreview}

Our work contributes to a long theoretical literature on the tendency to choose confirmatory sources and the polarization of beliefs. A decision maker receiving a binary signal only wants to hear that their prior expectation is wrong if the signal is strong enough to outweigh their prior. As such, a signal that tends to confirm the receiver's prior (but is very likely correct when it disagrees) can be preferable to an unbiased signal \citep{calvert1985value}. \citet{suen2004self} shows that using such biased sources over time, while rational, can lead to polarization in beliefs. \citet{nimark2019inattention} extend this work by incorporating rational inattention. We show that even if the full spectrum of signal realizations is available, it can be optimal to focus attention on a subset.\footnote{See also \citet{che2019optimal}, who study a stopping problem with two information sources, one biased towards each state, but no variation in quality. \citet{hu2021rational} extend this model to show that news consumers prefer to pay scarce attention to other consumers with similar beliefs and preferences, creating echo chambers when consumers learn from one another as well as biased primary sources.}

Another strand of the literature features heterogeneity in the preferences or prior beliefs of those one is learning from. Heterogeneity in prior beliefs can lead to information segregation between groups \citep{sethi2016communication}. Similarly, it may be preferable to learn from the opinions of those with similar preferences, if someone's opinion reflects both what they know and what they want \citep{williams2021echo}. In models of persuasion, echo chambers can offer those with similar beliefs \citep{meng2021learning} or preferences \citep{jann2018echo} an opportunity to credibly share information they otherwise might have misrepresented in an attempt to influence others.\footnote{See also \cite{giovanniello2021echo}.} By contrast, the driving mechanism behind the confirmation bias in our paper is heterogeneity in accuracy.

\citet{benoit2019populations} show that rational agents can update their beliefs in opposite directions in response to the same signal if the signal is ``equivocal''---if its likelihood depends on some ancillary issue, about which they have different beliefs. Disagreement can also persist when people use different models \citep{mailath2019learning}, face uncertainty about signal distributions \citep{acemoglu2016fragility}, or have higher-order misperceptions about each other's biases \citep{murooka2025bayesian}. 

The antecedent model closest to ours is that of \citet{gentzkow2006media}, who show that a Bayesian learning about a binary state will infer that news sources are more accurate if their reports have agreed with her beliefs in the past. If news producers value consumer perceptions of their accuracy, they will bias their reporting towards consumer beliefs to build trust.\footnote{See also \citet{innocenti2022can}, who shows that having a plurality of experts with opposite biases can lead to echo chambers that hinder learning.} Our continuous state space is a natural way to think about similarity between different states,\footnote{For example, if the state of the world represents the optimal minimum wage policy, it is intuitive to think that \$16 is more similar to \$15 than \$6.} and we show that the key intuition from \citet{gentzkow2006media} carries over: reports close to the prior expectation are of higher expected accuracy (Proposition \ref{prop:hwithinr}).\footnote{This is also true in \citet{gentzkow2025ideological}, whose model features normally distributed signals, discussed further below.} 

The remainder of our results require the richer state space, and have no analogues in prior work. Our main mechanism is a novel trade-off between credibility and usefulness that affects learning from others whenever there is uncertainty about the accuracy of their information. Since consumers in \citet{gentzkow2006media} and other binary-state models are choosing between just two actions, confirmatory reports are of no immediate use: they cannot change the consumer's action, and so she would never want to rule out opposing views entirely. By contrast, with a richer signal and state space we are able to show that confirmatory reports can be the most valuable, the optimal response can exhibit counterintuitive non-monotonicity, and it can be beneficial to ignore contradictory signals entirely. Importantly, the welfare implications of this approach differ from those of prior work. The biased reporting in \citet{gentzkow2006media} takes the form of garbling: randomly mis-reporting when the state of the world does not match the consumer's prior. This pandering makes the consumer worse off---a sharp contrast to our finding that a curated set of facts can both build trust and inform. Finally, we also apply our results to learning from peers.

Our setup does not rely on biased signals (though we allow for them), as our agent can focus on confirmatory reports by sampling unbiased signals with realizations close to her prior expectation. The results do not depend on discontinuities created by binary signals, or the boundedness of their support. And we do not require reputation built over time---even in our one-shot setup, credibility instantly accrues to viewpoints similar to the agent's own.

We study a rational agent who simply wants to know the state of the world. By contrast, another strand of related literature departs from fully rational Bayesian updating of beliefs.\footnote{\citet{rabin1999first} study an agent who occasionally misinterprets unbiased signals when they disagree with the prior belief. \citet{gentzkow2025ideological} show that a slight bias in what is thought to be unbiased direct information can lead to polarization in trust of secondary sources. In their model an agent observes signals from all sources; we complement their findings by studying how, even without biased signals, one's choice of sources interacts with mistrust to affect the accuracy and responsiveness of beliefs. \citet{loh2019dimensionality} show that if boundedly rational agents update marginal beliefs but not a joint distribution, beliefs can diverge asymptotically in response to common information. \citet{golman2017avoidance} review work on avoiding free information; in our model, information has an opportunity cost. \citet{stone2019just} models affective polarization, in which players' distrust of others' selflessness grows; our model has no misalignment of incentives. \citet{demarzo2003persuasion} study a failure to account for repetition of information. \citet{bowen2021learning} show polarization can occur when people sharing information misperceive others' access to primary sources.}
However, experimental findings support the idea that selective exposure to similar views is motivated by seeking credible information sources under time or attention constraints. \citet{fischer2005selective} find that limiting the number of information sources participants can review increases their tendency to select information that supports their prior views. \citet{metzger2020cognitive} find that news articles that support participants' prior attitudes are viewed as more credible. \citet{zhang2021partisan}	find evidence that partisans don't just dislike listening to those who are politically opposed, they sincerely believe them to be less informed. Even when selective exposure is associated with inaccurate beliefs (as \citet{cookson2020echo} find of stock investors), this may be explained by those in echo chambers simply having lower budgets for information or attention. Relative to work that does seem to find irrational levels of confirmation bias (e.g. \cite{hill2017learning}\footnote{\citet{hill2017learning} finds that subjects learning about political facts do update beliefs, but more slowly than Bayes' rule would suggest; in their experiment signal quality is known, unlike our setup.}), fully rational models such as ours may be thought of as a relevant baseline on top of which psychological biases may play a role. Further, our model may explain why such biases arose---as a heuristic approximating a rational mistrust of those with opposing views.

Finally, our paper can also be related to the literature on network formation. In social networks connected individuals often share traits (homophily), including political affiliation: friends tend to agree on politics. \citet{tenev2019} demonstrate this empirically, and show that networks of friendship or cooperation may form amongst politically aligned people as a social reward to motivate costly voting. We provide a complementary explanation for homophily in beliefs, beyond heterogeneity in preferences: people befriend those whose opinions they trust. And amongst similarly knowledgeable peers with different prior expectations, our model generates a network of communication where trust is reciprocal (Corollary \ref{cor:netform}).

\section{Model}\label{sec:model}
Consider an agent choosing an action $a\in \R$. The state of the world $\omega\in\mathbb{R}$ is unknown, and the agent has a prior belief $\mu_0(\omega)$ which has finite mean $\omega_0$ and finite variance $\sigma_0^2$. The state could be economic growth over the next year, the current rate of global warming, or the number of hospitalizations caused by a new virus, to give a few examples. The agent wants to match her action to the state of the world: her utility is $u(a\vert\omega)=-(a-\omega)^2$. 

There exists a continuum of signal sources. These might be peers or news organizations---both applications are discussed further below. A fraction $h$ are high quality and produce reports (signal realizations) distributed $f_H(s|\omega)$, while the remaining fraction $1-h$ are low-type, with signal realizations distributed $f_L(s|\omega)$. High-type signals are unbiased (the mean of $f_H\left(s\vert\omega\right)$ is $\omega$, the true state), though low-type signals need not be.\footnote{The assumption that high-type signals are unbiased is not important, since the agent could simply de-mean the observed signals if the more informative type had a large, known bias. However, our main results are robust to a fixed high-type bias, and moreover we are unaware of any realistic example where more informative sources are not also more accurate.} Both $f_H$ and $f_L$ are strictly positive everywhere, bounded, and continuous. We denote the variances of the high- and low-type signal distributions $\sigma_H^2$ and $\sigma_L^2$, and assume they are finite, with $\sigma_H^2\leq\sigma_L^2$. Define $f\left(s\vert\omega\right) \equiv hf_H\left(s\vert\omega\right) + \left(1-h\right)f_L\left(s\vert\omega\right)$ to be the total probability density of signal $s$ given state $\omega$. 

A signal realization $s$ is a report that the state $\omega$ is equal to $s$, and $\left(s-\omega\right)^2$ thus measures how inaccurate the report is. The key difference between the two types of signals is that low-type signals are less accurate: $\E\left[\left(s_L-\omega\right)^2\right] \geq \E\left[\left(s_H-\omega\right)^2\right]$.\footnote{This follows immediately from low type sources having higher variance and possible bias.} This may be because low-type sources make worse estimates, or because they make more mistakes in recollection or reporting, or because they are more biased or deliberately deceitful. We also assume that low-type signals are (Blackwell) less informative than high type signals, and that their informativeness to the agent vanishes as their variance increases.\footnote{Specifically, the expected utility of an agent with prior $\mu_0$ who observes one signal known to be low-type goes to $-\sigma_0^2$ (the value of observing no signals) as $\sigma_L^2\rightarrow\infty$.} This should seem natural given that they are less accurate, but rules out the possibility that the agent somehow knows the low type sources well enough to unravel their mistakes or misreporting so as to end up with better inferences than can be made from the more accurate high-type signals. While inaccuracy is not the only way one signal can be less informative than another,\footnote{For example, low-type signals might instead be low-variance but completely unrelated to the true state, in which case sampling close to the agent's prior expectation would not be an effective way of screening out low type signals. However, note that the agent would not be led too far astray by such signals.}  heterogeneity in accuracy is a feature of many important real-world learning environments---see e.g. \citet{d2012some} or \citet{allcott2017social}. We assume that the signal noise is orthogonal to the agent's prior noise,\footnote{If the agent realized she were reading a news report that cited the same primary source she had used to form her prior, for example, she would want to disregard it in favor of new information.} and that the cross-sectional distribution of reports in state $\omega$ is $f\left(s\vert\omega\right)$, as when signal noise is independent across the continuum of sources.\footnote{This does not require low-type errors to be unrelated---$f_L$ may be biased or concentrated away from the state, so that low types are wrong in the same way. It does rule out realization-specific common shocks that would make the mix of types within the sampling radius random, since the sampling law in Eq. \ref{eq:truncatedSignal} takes that mix as given by $f$.}

We assume that low-type signals are from a family of densities parameterized by variance that satisfies the following regularity condition: for any bounded interval $\left[\underline{s},\bar{s}\right]$ and state $\omega$,
\begin{equation}\label{eq:fzero}
	\sigma_L^2\rightarrow\infty \Rightarrow \sup_{\left[\underline{s},\bar{s}\right]} f_L\left(s\vert\omega\right) \rightarrow 0.
\end{equation}
Equation \ref{eq:fzero} says that the low-type density converges to zero on any interval as its variance goes to infinity. In other words, increasing the variance parameter does not simply make the tails longer, but can make any given region of the support arbitrarily unlikely. This property should not seem unnaturally restrictive---Eq. \ref{eq:fzero} holds for the normal distribution and the exponential distribution, among others.\footnote{Furthermore, even if the low-type signal distribution does not satisfy Eq. \ref{eq:fzero}, its tails always will (otherwise the distribution would be improper). That is, there will always be a fixed radius around the true state outside which $f_L\left(s\vert\omega\right)\rightarrow0$ as $\sigma_L\rightarrow \infty$. If the tails are less informative than the remainder of the signal distribution (low and high type combined), then we can relabel the tails the low type signals and proceed.}

Before choosing an action, the agent samples one signal. Its type (accuracy) cannot be credibly conveyed to the agent, but the agent can choose to focus on confirmatory reports and sample only from the sub-population of reports within a radius $r\in\bar\R_+$ of her prior mean $\omega_0$. Given a chosen range $r$, let $T_r$ denote the truncation operation. The distribution of the report sampled is then: 
\equa\label{eq:truncatedSignal}
T_rf(s|\omega)=\left\{
\begin{array}{c l}
	\frac{f(s|\omega)}{F(\omega_0+r|\omega)-F(\omega_0-r|\omega)} &; s\in \left[\omega_0-r,\omega_0+r\right]\\
	0&; \mbox{ otherwise }
\end{array}
\right.\fequa

where $F(r|\omega)=\int_{-\infty}^rf(s|\omega)ds$ is the cumulative distribution function of $f(\cdot)$. For $r=0$, let $T_0 f\left(s\vert\omega\right)$ be a point mass at $\omega_0$.

The timing is as follows.
\begin{enumerate}
	\item The state of the world $\omega$ is drawn from the agent's prior belief distribution $\mu_0$, which has mean $\omega_0$ and variance $\sigma^2_0$.
	\item The agent chooses a radius $r$ to sample within.
	\item The agent observes a signal realization $s$ drawn from the sub-population of reports within the sampling radius, distributed $T_r f\left(s\vert\omega\right)$.
	\item The agent takes an action $a\in\mathbb{R}$, and receives a payoff $u(a\vert\omega) = -( a - \omega)^2$.
\end{enumerate}

The agent can sample the whole space of signal realizations by choosing $r=\infty$.\footnote{In this case, let $T_\infty$ simply be the identity operator.} We call this strategy ``un-censored'' sampling, since the agent updates her beliefs based on the un-truncated realization of $f(s|\omega)$. However, while increasing the sampling radius $r$ is beneficial whenever the state of the world is far from the prior mean $\omega_0$, if low-type sources are sufficiently inaccurate, reports within a given radius of $\omega_0$ are likely to come from high-quality sources (see Proposition \ref{prop:hwithinr}). Therefore, restricting the sampling radius may maximize the informativeness of the signal. Crucially, the way to think about censoring in this model is not that the agent receives and discards reports outside radius $r$. After all, a defining feature of an echo chamber is that one does not hear what is outside it. Rather, by censoring the agent guarantees that the report drawn will not be outside the sampling radius, drawing from the sub-population within it and conditioning her belief update on her selection.\footnote{If signals are biased, censoring to an asymmetric interval around $\omega_0$ might (if allowed) yield higher utility than the symmetric radius $(\omega_0-r,\omega_0+r)$ required by our setup. Nonetheless, for simplicity we only allow a symmetric interval.} 
 
\section{Main results}\label{sec:results}

Given the problem described above, let $a^\star\left(s,r\right)$ be the agent's optimal action after having chosen sampling radius $r$ and then received report $s$. We denote the agent's optimal choice of sampling radius $r^\star$. The agent's posterior belief on the  state of the world after receiving signal  $s$ is:
\begin{eqnarray}\label{eq:posterior}
	\mu_1\left(\omega\vert s,r\right) 
	&=& 
	\frac{T_rf(s|\omega)\mu_0(\omega)}{\int T_rf(s|\omega^\prime)\mu_0(\omega^\prime)d\omega^\prime}.
\end{eqnarray}

Since the agent maximizes the expected value of $u(a\vert\omega)=-(a-\omega)^2$, the optimal action given a sampling strategy equals the posterior expected state. Another way of thinking about it is that since utility is quadratic loss, maximizing utility is equivalent to minimizing the expected variance of the posterior. Given a report and the sampling strategy, the optimal action is:\footnote{Throughout, we assume that this posterior expectation exists and is bounded over signals within any sampling interval. A sufficient condition is that for every finite $r$, $\int\left(1+\vert\omega\vert\right)\mu_0\left(\omega\right)\sup_{\vert s-\omega_0\vert\leq r}T_r f\left(s\vert\omega\right)d\omega<\infty$. This mild joint restriction on the prior and signal distributions holds in the normal model of Section \ref{sec:normaldist}, and can fail only if, for states far from $\omega_0$, some point within the sampling region is far more likely than the region as a whole.\label{fn:integrability}}
	
\begin{equation}\label{eq:optimala}
	a^\star(s,r)\;=\;\int \omega \mu_1\left(\omega\vert s,r\right) d\omega.
\end{equation} 

Assuming the agent plays optimally, we will write the expected loss of choosing sampling radius $r$ when the true state is $\omega$ as
\begin{equation}\label{eq:V}
	V\left(\omega,r\right) \equiv \int \left(a^\star\left(s,r\right)-\omega\right)^2 T_r f\left(s\vert\omega\right)ds.
\end{equation}

The agent's expected utility from choosing radius $r$ is then
\begin{equation}\label{eq:objective}
	EU(r) \;\equiv\; -\int \mu_{0}(\omega)\, V\left(\omega,r\right) d\omega,
\end{equation}

and the agent's objective in choosing $r\in \bar\R_+$ is to maximize $EU(r)$. In this setting, Lemma \ref{lem:brevis} shows that a report's value is its ability to change the agent's beliefs, and induce a posterior expectation $a^\star$ that differs from the prior, $\omega_0$.

\begin{lemma}\label{lem:brevis}
	For any radius $r \in \bar{\mathbb{R}}_{+}$, the expected utility is
	\begin{equation}
		EU(r) \;=\; -\sigma_{0}^{2} \;+\; \mathbb{E} \big[\big(a^{*}(s,r) - \omega_{0}\big)^{2}\big].
	\end{equation}
\end{lemma}

The agent will thus choose the censoring policy $r$ that maximizes how much she expects to learn. To study this choice, we define the agent's posterior expectations conditional on each possible source type for the observed signal. Let 
\begin{equation}\label{eq:optimalagiventype}
	a^\star_{q}(s,r)\;=\;\int\omega\mu_1(\omega|s,r,q)d\omega
\end{equation}
for source quality $q\in\{H,L\}$, where $$\mu_1(\omega|s,r,q) \;=\; \frac{ f_q(s|\omega)\cdot\mathbf{1}\{s\in\left(\omega_0-r,\omega_0+r\right)\}\frac{\mu_0(\omega)}{P_r\left(\omega\right)}}{\int f_q\left(s\vert\omega^\prime\right)\frac{\mu_0\left(\omega^\prime\right)}{P_r\left(\omega^\prime\right)}d\omega^\prime}$$ and $P_r\left(\omega\right) \equiv F\left(\omega_0+r\vert\omega\right)-F\left(\omega_0-r\vert\omega\right)$ is the chance of a signal within the sampling radius $r$ of the agent's expectation $\omega_0$.

It will be useful to note that the optimal action given a report $s$ and censoring radius $r$ is separable in the following sense.

\begin{lemma}\label{lem:weighted_a}
	For any censoring radius $r$, the optimal action is a convex combination of the expected posteriors conditional on the observed signal's source type, $a^\star_H(s,r)$ and $a^\star_L(s,r)$:
	$$ a^\star(s,r)=P(H|s,r)a^\star_H(s,r)+P(L|s,r)a^\star_L(s,r)$$
	where $P(H\vert s,r)\equiv\frac{h\int f_H(s\vert\omega^\prime)\mu_0(\omega^\prime)/P_r(\omega^\prime)\,d\omega^\prime}{\int f(s\vert\omega^\prime)\mu_0(\omega^\prime)/P_r(\omega^\prime)\,d\omega^\prime}$ is the posterior probability of the signal being of high quality given the report received and the censoring policy, and $P(L|s,r)=1-P(H|s,r)$ is the chance of a low-quality source.
\end{lemma}

We next show that the expected utility of playing uncensored is (unsurprisingly) lower than it would be if the source type were known.

\begin{lemma}\label{lem:Hbetter}
	The expected utility of not censoring (and playing optimally thereafter), $EU\left(r=\infty\right)$, is no greater than $hEU_H\left(r=\infty\right) + \left(1-h\right)EU_L\left(r=\infty\right)$, where $EU_H\left(r=\infty\right)$ and $EU_L\left(r=\infty\right)$ are the expected utilities of playing optimally without censoring with high and low types only.
\end{lemma}

With these results in hand, we can proceed to the main results. Proposition \ref{prop:hwithinr} shows that if low-type signals are inaccurate enough, reports close to the prior expectation are likely to come from high types.

\begin{prop}\label{prop:hwithinr}
	Fix prior beliefs $\mu_0$ and high-type signal distribution $f_H$ (with variances $\sigma_0^2$ and $\sigma_H^2$), the fraction of high types $h$, and a finite radius $r$. Let $P_q\left(\omega,r\right)\equiv\int_{\omega_0-r}^{\omega_0+r}f_q\left(s\vert\omega\right)ds$ be the chance that in state $\omega$, a type-$q$ report falls within the sampling radius. Then as low-type signal variance $\sigma_L^2$ goes to infinity, the ex-ante probability that the sampled report comes from a high-type source goes to one:
	$$\lim_{\sigma_L^2 \to \infty} \int \mu_0\left(\omega\right) \frac{hP_H\left(\omega,r\right)}{hP_H\left(\omega,r\right)+\left(1-h\right)P_L\left(\omega,r\right)}\, d\omega = 1.$$
\end{prop}

Since the main driver of this result is that some sources are more accurate than others---a ubiquitous feature of learning environments---this result is widely applicable. \citet{gentzkow2005wp}\footnote{The supplemental appendix to this working paper preceding \citet{gentzkow2006media} provides detailed formal results.} show that with a binary state space ``a source is judged to be of higher quality when its reports are more consistent with the agent’s prior beliefs,'' to which our Proposition \ref{prop:hwithinr} provides an analogue for a continuous state space. 

Despite our one-shot setup (which distinguishes our model from repeated games such as \cite{gentzkow2006media} and \cite{williams2021echo}), this expectation of quality is immediately useful to the agent. Specifically, confirmation bias can aid inference given the constraints. If low types are inaccurate enough, sampling only reports close to one's prior expectation can make posterior beliefs more accurate and improve welfare (Proposition \ref{prop:finiter}.\ref{prop:finitera}). Notably, only relative precision matters: censoring can be optimal because low types are sufficiently inaccurate, or because high types and the agent's prior are sufficiently precise (Proposition \ref{prop:finiter}.\ref{prop:finiterb}).

\begin{prop}\label{prop:finiter}\leavevmode
	\begin{enumerate}
		\item\label{prop:finitera} Fix the fraction of high-type sources $h\in\left(0,1\right)$, high-type noise $f_H$, and a prior $\mu_0$. Then there exists a threshold $\nu^\star>\sigma_H^2$ such that if the low-type signal variance $\sigma_L^2\geq \nu^\star$, censoring yields strictly higher expected utility than not censoring.
		\item\label{prop:finiterb} Alternatively, fix $h\in\left(0,1\right)$, $f_H$, $f_L$, and a prior $\mu_0$ which has (without loss) mean $\omega_0=0$. Assume that $f_L$ belongs to a location--scale family,\footnote{That is, $f_L\left(s\vert\omega\right)$ depends on $(s,\omega)$ only through $s-\omega$, and for every $\nu\geq 1$ the family includes $\frac{1}{\nu}f_L \left(\frac{s}{\nu}\vert\frac{\omega}{\nu}\right)$.} and whose informativeness vanishes as its variance grows, as assumed in Section \ref{sec:model}.
		For $\nu\ge1$, define
		\[
		\tilde\mu_0^\nu\left(\omega\right)\equiv\nu\,\mu_0\left(\nu\omega\right),
		\qquad
		\tilde f_H^\nu\left(s\vert\omega\right)\equiv\nu\,f_H\left(\nu s\vert\nu\omega\right).
		\]
		Then there exists $\nu^{\star\star}<\infty$ such that for all $\nu\ge\nu^{\star\star}$, an agent with prior $\tilde\mu_0^\nu$, high-type density $\tilde f_H^\nu$, and low-type density $f_L$ finds it optimal to censor.
		That is, for any fixed low-type variance, censoring is optimal if the prior and high-type signals are precise enough.
	\end{enumerate}
\end{prop}

The essence of the proof (in Appendix \ref{app:proofs}) is that if low types are inaccurate enough relative to the prior and high types, the agent can set a sampling radius $r$ such that almost all reports from high types are within $r$ and almost all reports from low types are outside $r$. Censoring thus approximates the expected utility of receiving only high-type signals, which is preferable to receiving both types (Lemma \ref{lem:Hbetter}). Low-type signal inaccuracy need not be extreme for our results to hold---censoring can still be optimal even when low-type signals are just as accurate as the agent's own prior expectation.\footnote{Both Propositions \ref{prop:hwithinr} and \ref{prop:finiter} do, however, make use of the lack of a bound on inaccuracy, which would not be the case if the state/signal space were restricted to a finite interval.} Nor is the near-complete filtering of low types afforded by Eq. \ref{eq:fzero} essential---see the example in Figure \ref{fig:sigdist} below. Also note that the proof of Proposition \ref{prop:finiter} does not require a symmetric sampling interval, so it is robust to other sampling strategies as well, e.g. intervals not perfectly centered around the agent's expectation. Furthermore, Corollary \ref{cor:leaky} (in Appendix \ref{app:proofs}) shows that the ability to perfectly exclude certain signal realizations is not essential---our results still hold if the echo chamber is ``leaky,'' occasionally providing reports from outside the sampling radius. 

There are two distinct applications of this model: learning from peers, and learning from news organizations. As a model of learning from peers, the sampling radius $r$ can be thought of as the range of views the agent is willing to entertain from those she consults before making a decision. On social media, this might mean following accounts whose profiles or reputations indicate similar views, or using ``block lists'' to filter out opposing opinions. Offline, it might entail joining a club of like-minded people, avoiding a new neighbor based on their bumper stickers, or even choosing where to live based on how residents tend to vote (as \cite{mccartney2024political} find). Having selected a group of peers based on some indication that their views are sufficiently similar, the agent finds out what one of them says the state is (e.g. by scrolling through social media, or asking a neighbor) before making her decision. 

Our model can alternatively be applied to a consumer choosing a news organization, as follows. There exist many sources (experts, scientific studies, etc.) that a news organization might cite, some more biased or inaccurate than others. The agent chooses a news organization known to only publish reports that fall within a certain range of the political spectrum and pays no attention to others, the budget constraint on signals leading to a form of rational inattention.\footnote{\citet{hu2021rational} study a similar rational inattention problem. Since their setup involves a binary state, only signals that challenge the agent's prior are useful, and there is no uncertainty about accuracy.} The news organization's editor plays a crucial intermediary role here, by only providing reports from the subset within a certain range such that others never reach the agent. This may be because the editor prioritizes certain pieces of evidence, to promote an agenda, or (as in \cite{gentzkow2006media}) to build trust---it is not costless to digest and summarize all the evidence about a particular topic. Alternatively, the editors themselves may not be as familiar with other evidence: writers may only pitch them stories matching the reputation of the organization, and editors may only develop relationships with expert sources that tend to confirm their views.

In both applications, we model the agent's signal extraction problem, but not peers' or news organizations' choice of what to report. This lets our results speak to a wide range of contexts in which one source type reports more accurately and informatively than another. The agent's selective exposure to confirmatory sources (a form of confirmation bias) results in an ``information silo'' or ``echo chamber'' as a rational response to realistic frictions: limited sampling capacity and uncertainty about the quality of signals. In both applications, the agent is aware that views outside her echo chamber exist, but given limited time or attention she may not be interested in hearing them.

\section{Normal prior and signals}\label{sec:normaldist}
Propositions \ref{prop:hwithinr} and \ref{prop:finiter} show that censoring can be optimal, but do not directly illustrate the main factors that determine the optimal sampling radius. To study the agent's choice of $r$ at the margin, this section imposes additional regularity assumptions: we assume the prior distribution $\mu_0$ and signal distributions $f_L$ and $f_H$ are normally distributed, with $\sigma_H^2<\sigma_L^2$ and unbiased low-type signals. This regularity permits sharper statements about the agent's choice of $r$. We formalize a novel tradeoff between the credibility of views close to one's own and the usefulness of those further away, and characterize the optimal sampling radius. 

\begin{figure}[H]\caption{Expected density of reports by source type\label{fig:sigdist}}
	\centering	
	\begin{minipage}{.8\linewidth}
		\includegraphics[scale=0.8]{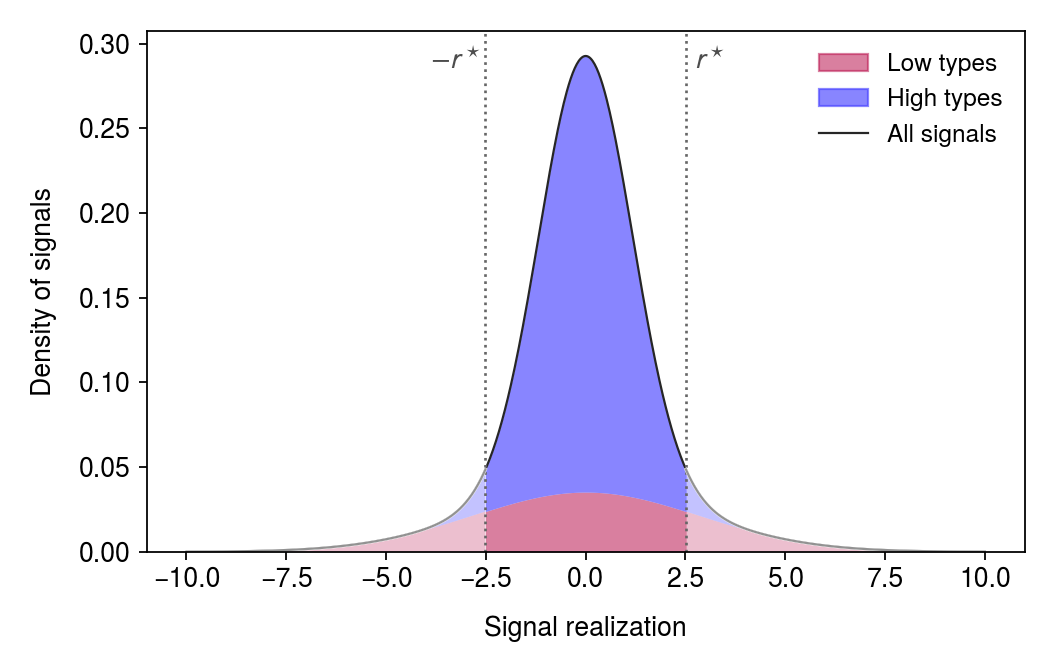}
	\end{minipage}
\end{figure}

We will illustrate the results in this section using the following example. Assume the variances of the prior and signals are $\sigma_0^2=1$, $\sigma_H^2=0.35$, and $\sigma_L^2=7$. The prior mean $\omega_0$ is zero and a quarter of the sources are low-type ($h=0.75$). Figure \ref{fig:sigdist} shows the prior expected distribution of reports (that is, marginalizing over all possible states). 

With normal signals these ex-ante distributions are also normal, with mean $\omega_0$ and variance $\tilde\sigma_q^2 \equiv \sigma_0^{2}+\sigma_q^{2}$ for $q\in\{ L,H\}$, so that $\tilde\sigma_H^2<\tilde \sigma_L^2$. Reports near the prior expectation are thus more likely than distant ones to come from high-type sources (whereas Proposition \ref{prop:hwithinr} required potentially extreme low-type variance to reach a similar conclusion).

With normal signals, standard results in Bayesian inference yield that the optimal action given source quality $a_q^*(s, \infty)$ is a linear convex combination of the prior expectation and the received signal. The weights are given by the relative accuracy of the prior and of the signal:
\begin{equation}\label{eq:normalPolicy}
	a_q^*(s, \infty) = \frac{\sigma_q^2}{\sigma_q^2+\sigma_0^2}\omega_0+\frac{\sigma_0^2}{\sigma_q^2+\sigma_0^2}s \: \: \:\text{   for } q\in\{L,H\}.
\end{equation}

This linearity between the signal and the action is broken when the agent chooses a finite sampling radius $r<\infty$, but the un-censored benchmark will nonetheless be useful for subsequent results such as the credibility-usefulness tradeoff (Corollary \ref{cor:creduseful}).

While the optimal action conditional on the source type is not linear, it is monotonically increasing.

\begin{lemma}\label{lem:sincreasing}
	Assume $\mu_0$, $f_L$, and $f_H$ are normally distributed, and $r < \infty$. Then the optimal action $a_q^\star\left(s,r\right)$ is strictly increasing in the signal realization $s$ for any source type $q\in\{ H,L\}$.
\end{lemma}

Lemma \ref{lem:sincreasing} does not require normality, just the monotone likelihood ratio property (higher reports make higher states relatively more likely), which any strictly log-concave location family of signal densities satisfies. It holds for any prior, since the censoring factor in Eq. \ref{eq:optimalagiventype} is common to the posteriors being compared. 

Next, with no censoring, we see that the ex-post belief that a signal is high quality (which determines the weights on the source-specific policy functions) is decreasing in the distance between the prior expectation and the report. 

\begin{lemma}\label{lem:normalodds}
	If $f_H(s|\omega), f_L(s|\omega)$ and $\mu_0(\omega)$ are normal and the agent is not censoring ($r=\infty$), the odds of a high-type source given a report are decreasing in $|\omega_0 - s|$ and take the following form:
	$$\frac{\text{P} (H|s,r=\infty)}{\text{P} (L|s,r=\infty)}=\left(\frac{h}{1-h}\right)
	\sqrt{\frac{\sigma_L^2+\sigma_0^2}{\sigma_H^2+\sigma_0^2}}\cdot \text{\em{exp}}\left\{-\frac{( s-\omega_0)^2(\sigma_L^2-\sigma_H^2)}{2(\sigma_H^2+\sigma_0^2)(\sigma_L^2+\sigma_0^2)}\right\}.$$
	Therefore, the probability of a high-type source is also decreasing in $|\omega_0 - s|$.
\end{lemma}

Lemma \ref{lem:normalodds} says that the more the signal received disagrees with the agent's prior, the less likely it is perceived to be high-quality information---decreasing credibility. With the regularity properties of normal distributions, this is true everywhere in the signal space as long as $\sigma_L^2 > \sigma_H^2$ and the agent is not censoring. (Proposition \ref{prop:hwithinr} is similar in spirit, but requires the low-type signal noise $\sigma_L^2$ to be high enough, and reaches a weaker conclusion.) Note that this principle can be applied to other contexts as well, such as scientific experiments. If you step on a scale and it tells you that you weigh a tenth of what you expected, most likely you need a new scale.

While reports far from the agent's expectation are less credible, they are more useful, having more potential to correct a mistaken belief. With normal primitives and no censoring, the influence of these two competing forces on the agent's beliefs can be factored cleanly.

\begin{corollary}[credibility--usefulness tradeoff]\label{cor:creduseful}
	Assume $\mu_{0}$, $f_{H}$, and $f_{L}$ are normal with $\sigma_{H}^{2} < \sigma_{L}^{2}$, $h \in (0,1)$, and no censoring ($r = \infty$). Then the influence of a report on the agent's beliefs is its credibility $w\left(s\right)$ times its usefulness, which has magnitude $\vert s-\omega_0\vert$:
	\begin{equation}\label{eq:tradeoff}
		a^{*}(s, \infty) - \omega_{0} \;=\; w(s)\, (s - \omega_{0}),
		\qquad
		w(s) \;\equiv\; P(H|s)\, \frac{\sigma_{0}^{2}}{\sigma_{0}^{2} + \sigma_{H}^{2}} \;+\; P(L|s)\, \frac{\sigma_{0}^{2}}{\sigma_{0}^{2} + \sigma_{L}^{2}}.
	\end{equation}
	Credibility $w\left(s\right)$ is maximized at $s=\omega_0$ and strictly decreasing in usefulness $\vert s-\omega_0\vert$, with $\lim_{\vert s-\omega_0\vert\rightarrow\infty}w\left(s\right) = \frac{\sigma_0^2}{\sigma_0^2+\sigma_L^2}$. 
\end{corollary}

This is the clearest form of the novel tension between credibility and usefulness inherent to our problem. Reports close to the prior expectation are more credible (in proportion to $P(H|s)$) but less useful, since they do not lead the agent to change her beliefs much. Reports far from the prior expectation are useful if true, but less credible. Note that this tradeoff is hard to illustrate in the binary state/signal space used in much of the literature. 

Note also that the credibility half of the tradeoff is not special to normality. $P(H|s)$ is maximized at $s = \omega_{0}$ and decreasing in $|s - \omega_{0}|$ whenever the ratio of the two types' ex-ante signal densities (Figure \ref{fig:sigdist}) is, and it vanishes in the tails whenever the low type signal density has a heavier tail (see footnote \ref{fn:MLRP}). Normality is required only for the multiplicative factorization in Corollary \ref{cor:creduseful}. 

Combining Lemma \ref{lem:brevis} with Corollary \ref{cor:creduseful}, we can write the value of the un-censored policy in terms of credibility and usefulness:

\begin{equation}\label{eq:EUinfty}
	EU(r=\infty) \;=\; -\sigma_{0}^{2} \;+\; \int m(s)\, w(s)^{2}\, (s - \omega_{0})^{2}\, ds,
\end{equation}
where $m(s) \equiv h\, \mathcal{N}\!\big(s;\, \omega_{0},\, \sigma_{0}^{2} + \sigma_{H}^{2}\big) + (1 - h)\, \mathcal{N}\!\big(s;\, \omega_{0},\, \sigma_{0}^{2} + \sigma_{L}^{2}\big)$ is the density of reports under the prior (Figure \ref{fig:sigdist}), using $\mathcal{N}\left(\cdot;\mu,\sigma^2\right)$ to denote the normal density with mean $\mu$ and variance $\sigma^2$. The value of sampling adds up the credibility-weighted usefulness of each potential report, weighted by the chance of receiving it.

This tension is clear when $r=\infty$, as shown above, and also when $r=0$, where the agent is guaranteed to receive the most credible signal $s=\omega_0$ but learns nothing from it. The optimal radius lies between, and the tradeoff that characterizes it is expressed by Proposition \ref{prop:rstarnormal}.

\begin{prop}\label{prop:rstarnormal}
	Assume $\mu_{0}$, $f_{H}$, and $f_{L}$ are normal with $\sigma_{H}^{2} < \sigma_{L}^{2}$ and $h \in (0,1)$. Then $EU$ is continuously differentiable on $(0, \infty)$, and
	\begin{equation}\label{eq:EUprime}
		EU'(r) \;=\; 2\, m_{r}(\omega_{0} + r)\, \Big[\; \E\big(\, V(\omega, r)\ \big|\ s = \omega_{0} + r,\ r \,\big) \;-\; \Var\big(\, \omega\ \big|\ s = \omega_{0} + r,\ r \,\big) \;\Big],
	\end{equation}
	where $m_{r}(s) \equiv \int \mu_{0}(\omega)\, T_{r} f(s|\omega)\, d\omega$ is the density of the signal the agent hears (so $m_{\infty} = m$). As a result, any interior optimal radius $r^{\star} \in (0, \infty)$ satisfies
	\begin{equation}\label{eq:tradeoffatrstar}
		\big( a^{*}(\omega_{0} + r^{\star},\, r^{\star}) - \omega_{0} \big)^{2} \;=\; \E\big(\, (\omega - \omega_{0})^{2} - V(\omega, r^{\star})\ \big|\ s = \omega_{0} + r^{\star},\ r^{\star} \,\big).
	\end{equation}
	If the conditions of Proposition \ref{prop:finiter} hold, an interior optimum exists that satisfies Equation \ref{eq:tradeoffatrstar}.
\end{prop}

Equation \ref{eq:EUprime} shows that when considering expanding the sampling radius, the agent weighs two quantities. The term $\Var(\omega\,|\,s = \omega_{0} + r,\, r)$ is the residual uncertainty having drawn a signal exactly at the boundary of the range the agent is considering expanding. The term $\E\big(\, V(\omega, r)\ \big|\ s = \omega_{0} + r,\ r \,\big)$ is the expected loss that a draw from within the current sampling radius would have delivered, conditional again on a signal at the boundary (because the agent is interested in the comparison in states likely to produce a signal at the boundary). 

Where these quantities balance, Equation \ref{eq:tradeoffatrstar} holds, and characterizes the optimal radius.\footnote{Even with  all primitives normal, no closed form for $r^\star$ exists, since the integrals defining the optimal action (Eq. \ref{eq:optimalagiventype}) include normal CDF terms.} Each side is phrased as gains over the uninformed action $\omega_0$, and conditioned on the states likely to produce a signal at the boundary. The left hand side of the equation is the marginal view's squared influence---its usefulness weighted by its credibility. The right hand side is the value of a report from within the echo chamber. The agent expands the range of views worth considering until the most extreme admissible signal leaves her just as uncertain as the average report.

\begin{figure}[H]\caption{Expected utility as a function of censoring radius $r$\label{fig:utilityr}}
	\centering	
	\begin{minipage}{.8\linewidth}		
		\includegraphics[scale=0.8]{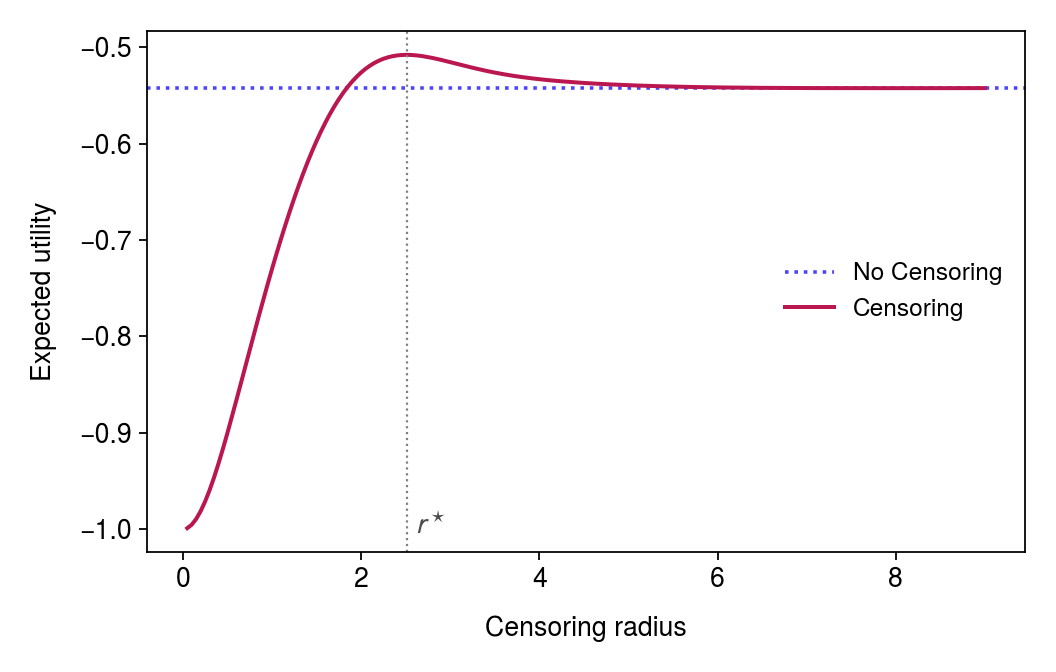}
	\end{minipage}
\end{figure}

Figure \ref{fig:utilityr} compares the expected utility of censoring to radius $r$ (solid line) to the expected utility of not censoring (dotted line), in all cases assuming the agent plays the optimal action given the signal received as per Equation \ref{eq:optimala}. At radius $r=0$, the agent is only sampling reports exactly equal to the prior expectation of 0, and not learning anything. Expected utility is thus simply equal to the negative prior variance, since utility is quadratic loss. Radius $r^\star\approx2.5$ maximizes the expected utility of censoring, at a level higher than the expected utility of not censoring. As $r$ goes to infinity, the expected utility with censoring approaches the expected utility with no censoring, as the radius of attention grows to encompass the full signal realization space.

\begin{figure}[H]\caption{Signal variance and correlation with the state as a function of censoring radius\label{fig:totalvar}}
	\centering
	\begin{minipage}{.8\linewidth}
		\includegraphics[clip, scale=0.8]{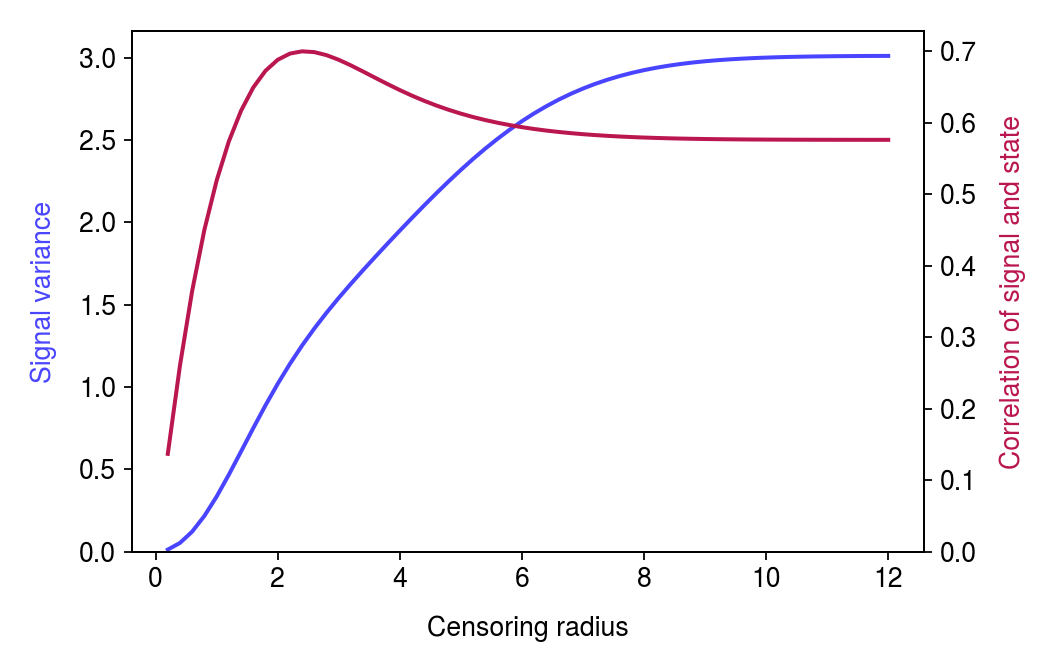}
	\end{minipage}
\end{figure}

Figure \ref{fig:totalvar} provides another way to see this tradeoff, by plotting the total variance of the signal and its correlation with the state of the world as the radius increases. Increasing $r$ raises the variance of the signal received, including valuable information about the state. But beyond the optimal radius $r^\star$, the correlation of the signal with the state drops as a low-type source becomes more likely.  

\begin{figure}[H]\caption{Optimal action as a function of report\label{fig:optimalaction}}
	\centering
	\begin{minipage}{.8\linewidth}
		\includegraphics[scale=0.8]{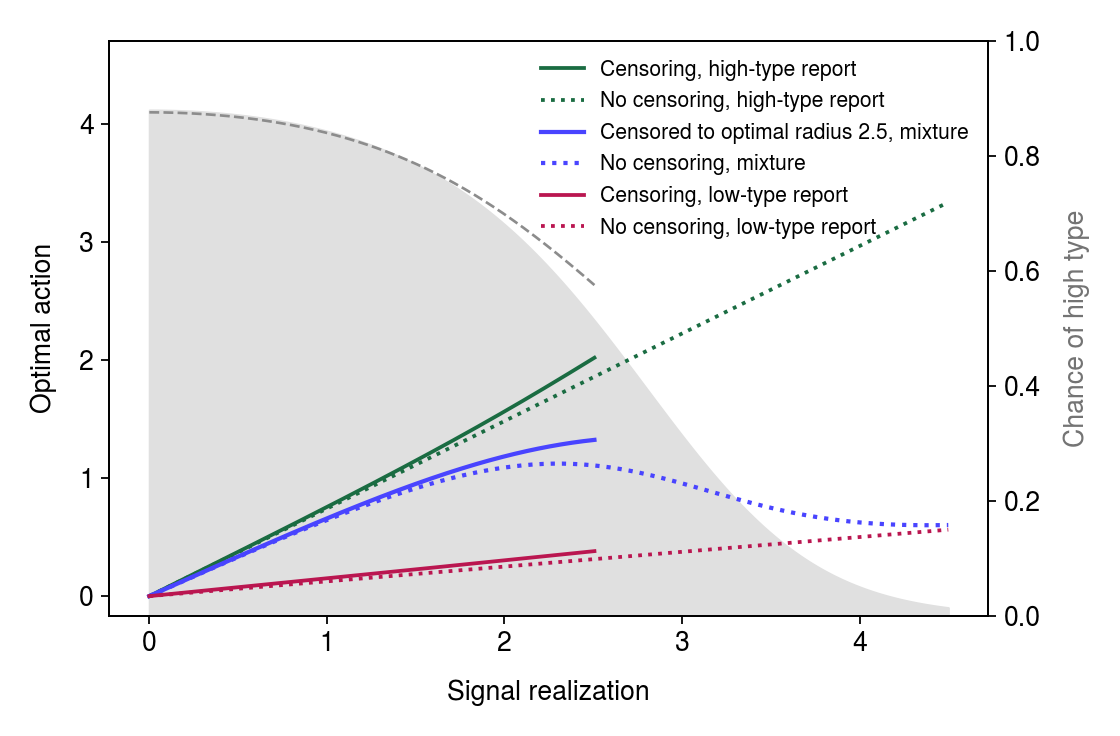}
	\end{minipage}
\end{figure}

Figure \ref{fig:optimalaction} illustrates these pieces put together for the aforementioned parameters. The dotted blue line in the middle is the optimal action (posterior expected state) with no censoring. Per Lemma \ref{lem:weighted_a}, this is a mixture of the optimal action conditional on a low-type source and the optimal action conditional on a high type, given by the dotted red and green lines respectively. The weight is determined by the chance of a high type for a given signal, given (when censoring to $r^\star$) by the grey dashed line (right axis); the grey shading represents the chance of a high type with no censoring (Lemma \ref{lem:normalodds}). Here higher signals are less likely to come from high types whether censoring or not, so the optimal action starts off closer to the upper high-types ray at the origin but approaches the lower low-types ray as the signal $s$ goes to infinity. 

The solid blue line in the middle shows the expected state if the agent samples only within the optimal radius $r^\star\approx 2.5$. This is again a weighted mixture of the solid red and green lines below and above, which represent the conditional-on-low-types and conditional-on-high-types optimal actions given censoring to $r^\star$. Note that the agent is more responsive to signals from the echo chamber in this example. 

Without censoring, the weight on the conditional-on-high-types response decreases as the signal gets further from the prior expectation (Lemma \ref{lem:normalodds}).\footnote{Without censoring, the odds of a high-type source given a report equal $h/(1-h)$ times the ratio of the two types' ex-ante report densities (Figure \ref{fig:sigdist}), so this monotonicity holds whenever that ratio is decreasing in $\vert s-\omega_0\vert$, and the chance of a high type vanishes in the tails whenever the low-type ex-ante density has heavier tails.\label{fn:MLRP}} With censoring, it is possible that the weight can increase toward the edge of the sampling interval, because in states far from the prior expectation, high-type reports that land within the sampling interval cluster at its nearest edge. In our example, however, it is decreasing throughout the optimal radius (Figure \ref{fig:optimalaction}).

An interesting property arising from signals with uncertain quality is that the optimal action (especially with no censoring) can exhibit non-monotonicity. Since signals further from the agent's prior expectation are less trustworthy, they can also be less persuasive. So there may be regions of the signal space in which the agent's posterior belief is decreasing in the signal received---a more extreme signal results in a less extreme action.\footnote{\citet{williams2021echo} derives a similar non-monotonicity owing to correlation in preferences.}

In the normal case, the non-monotonicity occurs even though both components that are averaged, $a_H^*(\cdot)$ and $a_L^*(\cdot)$, are monotonically increasing. The average is non-monotonic because the weights are not constant. For example in Figure \ref{fig:optimalaction}, a report of $s=2$ prompts a higher action than  $s=4$, because the agent knows that the latter almost surely came from a low-type source. In other words, a moderate view can elicit a greater response than an extreme one, because it is more trustworthy. Proposition \ref{prop:actionnonmon} gives conditions for this non-monotonicity to occur.

\begin{prop}\label{prop:actionnonmon}
	Assume that the prior $\mu_0$ and signals $f_L$ and $f_H$ are distributed normally. Fix $\sigma_0^2$, $\sigma_H^2$, and $h\in\left(0,1\right)$. Then there exists a threshold $\bar\sigma_L^2$ such that $\sigma_L^2>\bar\sigma_L^2$ implies that the optimal action without censoring $ a^\star(s,\infty)$ is not monotonic in $s$, and has a local maximum.
\end{prop}

The uncertainty about the quality of a signal is key for the optimal response to be non-monotonic. So long as $h$ is in the interior, and the difference in quality between high-types and low types is big enough, the transition from an optimal response close to the high-type-only response to one close to the low-types-only response will imply some non-monotonicity. In the normal case this gap is completely dictated by the difference in variances, but non-monotonic responses can also be expected in more general settings.

Finally, with normal signals, uncertainty about the signal's quality is not only sufficient for censoring to be optimal (Proposition \ref{prop:finiter}), but also a necessary condition. In the limit where all sources are one type, or the limit where high and low types are equally noisy, the optimal radius grows without bound. And at that limit, with no uncertainty about quality, uncensored sampling is uniquely optimal. 
\begin{prop}\label{prop:actionlin}
	Assume $f_L$, $f_H$, and $\mu_0$ are distributed normally, and let $\left(\sigma_{H,n}^2,\sigma_{L,n}^2,h_n\right)$ be a sequence of parameters converging to a limit $\left(\sigma_H^2,\sigma_L^2,h\right)$, with either $h\in\{0,1\}$ (all sources one type) or $\sigma_H^2=\sigma_L^2$ (both types equally noisy). Then the optimal radius satisfies $r^\star_n\rightarrow\infty$, and for every signal realization $s$ the optimal action $a^\star\left(s,r^\star_n\right)$ converges to the increasing linear function of $s$ in Eq. \ref{eq:normalPolicy}, $\frac{\sigma_q^2}{\sigma_q^2+\sigma_0^2}\omega_0+\frac{\sigma_0^2}{\sigma_q^2+\sigma_0^2}s$, where  $\sigma_q^2$ denotes the variance of the source type surviving at the limit.\footnote{ That is, $\sigma_q^2=\sigma_L^2$ if $h=0$, and $\sigma_q^2=\sigma_H^2$ otherwise.}
\end{prop}

The intuition of the proof is that if there is no uncertainty about the quality of information, the signal received is normal, so the posterior expected state is a linear-convex combination of the prior and the signal, and un-censored sampling attains the Bayes risk of that problem. A Bayesian version of the Cram\'er--Rao bound (the van Trees inequality) then shows that any finite sampling radius does strictly worse than not censoring.

\begin{figure}[H]\caption{Expected action as a function of state, by censoring radius\label{fig:expaction}}
	\centering
	\begin{minipage}{.8\linewidth}
		\includegraphics[scale=0.8]{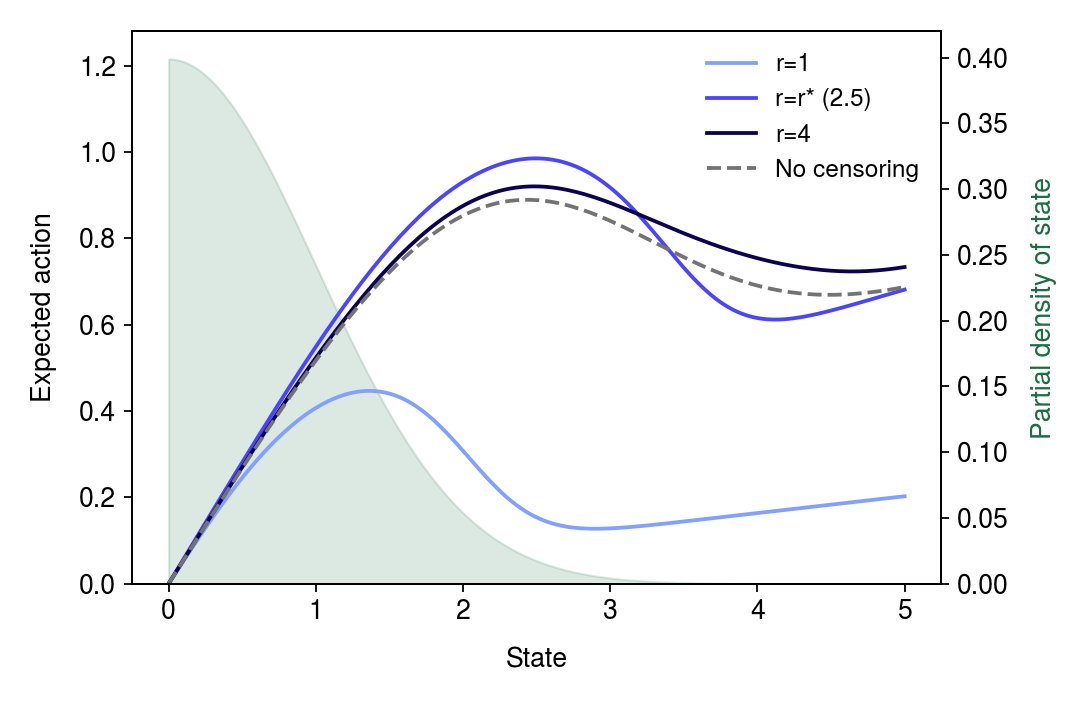}
	\end{minipage}
\end{figure}

Given these results, what is the best way to disabuse the misinformed? Allowing someone a signal from their echo chamber may move their beliefs further towards the truth than simply exposing them to views they don't trust. Figure \ref{fig:expaction} plots the agent's expected action as a function of the true state, for various censoring radii. Since the agent's prior expectation is zero, the further the true state is from zero the more wrong the agent's prior beliefs. While no radius dominates over all possible states, signals from the optimal radius $r^\star$ yield a greater expected response than un-censored signals for all but the most extreme states.\footnote{If the prior expectation is very wrong (in this case, more than ~3 standard deviations from the true state), then an un-censored signal is better ex post. But since the agent believes such a state is extremely unlikely, this is deemed a risk worth taking.} So when source quality is uncertain, confirmatory sampling may provide the best chance of correcting mistaken beliefs, since only trusted news is heeded. Note, however, that this result relies on the assumption that even within one's echo chamber, signals are more likely to be close to the true state than further away. 

\section{Results for populations of agents}\label{sec:pop}

Next, we apply our model to populations of agents. We assume that all agents are simply trying to learn the state, as above. Each agent can choose a distinct sampling radius, and will observe a report drawn from within it (which will almost certainly not be the same as the report drawn by a given other agent).

First, we show that selective exposure to confirmatory views moves the beliefs of a population closer to the truth on average. Confirmation bias is often assumed to increase the polarization of beliefs, but we show that when used to filter out inaccurate information it can instead reduce polarization, provided that initial beliefs are sufficiently dispersed to begin with.\footnote{\textcite{faia2021biases} provide experimental evidence that people's beliefs diverge after reading opposing articles, and more so if they chose the articles themselves based on the headlines. This is consistent with our model, which predicts that disparate signals can move beliefs in opposite directions, and that trusted news matching one's prior can be more persuasive. Since the two articles used in their experiment were not random draws but rather selected specifically to represent two opposing viewpoints, the overall effect of selective exposure to news cannot be inferred from their experiment. In fact, since the articles were specifically chosen to represent views not typically associated with their sources, it is possible that they actually fostered convergence of beliefs amongst readers outside the experimental setting, if readers chose what to read based on source rather than headlines.\label{fn:faia}}

\begin{corollary}\label{cor:polarization}
	Fix a population of $N$ agents indexed by $i$. Suppose nature draws the state, each agent draws an i.i.d. prior expectation conditional on the true state, and priors are calibrated. Then optimal sampling weakly reduces the mean squared distance of posterior expectations from the true state, $\E\big[\tfrac{1}{N}\sum_i \big(\overline{\mu_{1}^i}-\omega\big)^2\big]$, relative to uncensored sampling. The reduction is strict if, with positive probability, an agent's prior is one for which censoring is strictly optimal (as under the conditions of Prop. \ref{prop:finiter}).
\end{corollary}

Corollary \ref{cor:polarization} does not require agents to start with the same prior expectation, nor does it require each agent to choose the same sampling radius. Since this result shows that posterior accuracy increases with censoring, it is natural to wonder when this translates to reduced disagreement. Since agents' errors are independent conditional on the state, their mean squared error (the quantity in Corollary \ref{cor:polarization}) is twice the expected squared distance between two agents' posterior expectations (disagreement) plus twice their common squared error, $\E\big[\big(\E[\overline{\mu_{1}^i}-\omega\,\vert\,\omega]\big)^2\big]$. The corollary shows that the sum of these idiosyncratic and shared components of error is reduced by optimal censoring, but we can only say that disagreement is reduced if the idiosyncratic component in particular decreases.

When prior expectations are sufficiently diverse, it does. The shared error is small relative to total accuracy when each agent's own uncertainty is small relative to initial disagreement (the dispersion of prior expectations in the population). In this case, a reduction in posterior error translates (to a first order) to a reduction in disagreement.\footnote{A formal statement and proof of the limiting case for the normal model of Section \ref{sec:normaldist} is available on request.} To illustrate this, consider the example of Section \ref{sec:normaldist}. Suppose nature draws the state from a normal distribution and each agent's prior expectation is her calibrated posterior mean given a private normal signal of the state, so that every agent's prior has variance $\sigma_0^2=1$. Compared to not censoring at all, optimal censoring reduces the mean squared error of agents' posterior expectations from $0.54$ to $0.51$ (regardless of how dispersed prior expectations are). If the variance of prior expectations across agents equals $\sigma_0^2$, the expected variance of posterior expectations across agents falls from $0.40$ to $0.38$---a reduction in disagreement as beliefs converge towards the truth.\footnote{Larger initial disagreements have similar results: if the variance of prior expectations is $3\sigma_0^2$, posterior variance falls from $0.47$ to $0.44$.}

When agents start from similar beliefs, however, the shared error can absorb the entire gain in accuracy. In the same example but with a common prior, optimal censoring raises the expected variance of posterior expectations across agents slightly, from $0.243$ to $0.245$, even as it brings them closer to the truth. Intuitively, when everyone starts in the same place, all disagreement comes from the different reports agents happen to draw, and optimal censoring makes each report more credible and hence more persuasive. The same logic applies within a like-minded subgroup: its more moderate members change their beliefs most in response to incoming news, which can increase disagreement within the subgroup even while reducing it in the population as a whole.

While optimal selective exposure improves the accuracy of beliefs on average, it can make some agents---those with very inaccurate priors---worse off in expectation. Our ability to control who we follow on social media provides us with trusted opinions, but can also make those with inaccurate beliefs more receptive to misinformation.

\begin{corollary}\label{cor:ineq}
	Suppose that as the state gets further from the prior expectation, the expected loss from uncensored sampling remains a strict fraction of the loss from remaining uninformed:\footnote{There are various specific conditions for heavy-tailedness under which the posterior expectation will be unbounded and track the signal---see e.g. \citet{o2012bayesian}. Condition \ref{eq:ineqcond} itself implies that the posterior expectation without censoring is unbounded.}
	\begin{equation}\label{eq:ineqcond}
		\limsup_{\vert\omega-\omega_0\vert\rightarrow\infty} \frac{V\left(\omega,\infty\right)}{\left(\omega-\omega_0\right)^2}<1.
	\end{equation}
	Then conditional on having a prior expectation far enough from the true state, using any finite sampling radius (including the optimal one) is worse in expectation than uncensored sampling.
\end{corollary}

Note that Condition \ref{eq:ineqcond} is unchanged if the prior expectation in the denominator is replaced by any fixed action.\footnote{Taking fixed action $\bar{a}$, note that $\left(\omega-\bar{a}\right)^2/\left(\omega-\omega_0\right)^2\rightarrow 1$ as $\vert\omega\vert\rightarrow\infty$} It can thus compare the uncensored agent's expected loss to that of any agent whose actions are bounded---in particular, one sampling from a finite radius. The condition holds in the normal model of Section \ref{sec:normaldist}, and whenever heavy prior tails lead the uncensored posterior expectation to track the signal closely enough that the loss $V\left(\omega,\infty\right)$ remains bounded. The intuition is that a finite sampling radius bounds the set of actions the agent might take in response to a signal, whereas with uncensored sampling there is no bound to the action the agent would take in response to an extreme enough signal.

Second, our results can be applied to network formation amongst peers. Suppose each signal source is another agent facing a similar problem. Each agent has beliefs about the state of the world, though some (high types) have more accurate expectations than others. Each chooses a radius that defines a set of peers whose expectations are within a certain distance from hers, and then learns the expectation of one of these peers before taking her action, with payoffs as before.\footnote{If agents have no incentive to misreport their beliefs, this is equilibrium behavior. If they do, each agent's choice of whom to link with remains a best response to the reports induced by others' strategies, and our results apply so long as those reports retain the necessary assumed properties.}

Here the agents and their choices form a graph, in which each agent is a node, and a directed edge from $i$ to $j$ indicates that $j$ is included in the set of trusted peers that $i$ will sample from. In this context, censoring results in homophily in beliefs: links (edges) are more common between like-minded peers, a phenomenon observed in many real-world settings (e.g. \cite{tenev2019} and \cite{mccartney2024political}). 

In general, two agents' sampling radii will not coincide unless their priors do as well, though they may have large regions of overlap. (Experts with large social media followings do not typically follow all of their followers.) However, amongst similarly confident agents, all trust is mutual: the subgraph of optimal sampling is reciprocal.

\begin{corollary}\label{cor:netform}
	Consider a subset of agents whose prior beliefs are translations of one another, and differ only in their means. Assume that $f_H$ and $f_L$ depend on $s$ and $\omega$ only through $s-\omega$. If each of these agents chooses their optimal sampling radius then it is optimal for all links amongst them to be bilateral: $i$ is in $j$'s sampling radius iff $j$ is in $i$'s.\footnote{In the case of multiple optimal radii, this requires all agents in the subset to make the same selection.}  
\end{corollary}

\begin{proof}
	Since each agent's prior belief is the same (up to a constant), the same sampling radius is optimal for all. The result follows immediately.
\end{proof}

To be clear, this pairwise trust (``mutuals'' on social media) might not persist indefinitely, were the model extended to include multiple periods of learning and link formation. But this is realistic: people's beliefs change as they learn about the world \citep{sharot2023and}, as does the set of peers they trust. Also, note that Corollary \ref{cor:netform} implies that in settings of peer learning, no intermediary is necessary to filter out untrustworthy signals. Simply having a coordinating mechanism to meet up with like-minded people, be it explicit (joining a group that supports a political candidate) or incidental (moving to a city that tends to vote a certain way) can suffice. 

Like our main findings (Section \ref{sec:results}), these results make few demands on the distributions of the prior or signals. Simply having low-type sources that are comparatively inaccurate enough---even if they are rare---is enough to make signals close to one's prior expectation better quality in expectation. As such, focusing attention on views similar to one's own can yield the most accurate posterior beliefs given the constraints, reduce disagreement in a population, and generate networks of mutual trust. 

\section{Secondary learning}\label{sec:secondary}

We supplement our study of peer learning by studying whether confirmation bias in sampling will encourage or discourage similar behavior in others.\footnote{We are grateful to an anonymous referee for this fruitful suggestion.} Suppose that first, a set of agents we will call ``primary'' learn from signals exactly as described in Section \ref{sec:model}. Then, a secondary learner (or group thereof) learns from the primary learners by selecting a sampling radius around his own expectation and learning the expectation of one primary agent within that radius. 

In the context of peer learning, primary learners are those with access to (or the sophistication to interpret) primary sources of information, such as scientists who post on social media about developments in their field. Secondary learners are those who learn from them (e.g. non-specialists). In the context of learning from news organizations, we can think of primary learners as the news media who choose which sources to use in their reports, and secondary learners as consumers who learn from them. 

Suppose that some primary learners choose not to censor but others (perhaps those with more precise priors---see Proposition \ref{prop:finiter}.\ref{prop:finiterb}) do. Censoring optimally improves the accuracy of a primary learner's beliefs (Proposition \ref{prop:finiter}), and by Corollary \ref{cor:polarization} the posterior expectations of primary learners who do censor are less dispersed around the true state than they would be without it, so any existing gap in accuracy owing to the difference between the groups' prior precisions widens. From the perspective of a secondary learner, then, this results in the same kind of heterogeneity in accuracy as in the sources in Section \ref{sec:model}, with the primary learner's sampling choice here playing the role of source type.\footnote{To make the analogy work, we assume that the secondary learner does not observe which primary learners censored.}

Suppose further that, given the state, both groups' posterior expectations are centered on it.\footnote{With calibrated priors this holds approximately when each group's prior uncertainty is small relative to the variation in the state across the population (see Section \ref{sec:pop}).} Then the posterior expectations of censoring primary learners are distributed more closely around the state than those of non-censoring primary learners. A secondary learner expects the state to be near his own prior expectation, so his problem resembles that of Section \ref{sec:normaldist}: if the two groups' belief distributions are ordered as are the two signal distributions there (Lemma \ref{lem:normalodds}), a primary learner with beliefs close to the secondary learner's expectation is more likely to be one who censored. We do not establish this ordering formally, and note that it can fail if non-censoring learners' opinions cluster near a common prior center rather than around the state, in which case a secondary learner near that center would find nearby reports more likely to come from non-censoring learners.

Whether this makes it optimal for the secondary learner to censor depends, as in the main model, on the gap in accuracy between the posterior expectations of the two groups.\footnote{Proposition \ref{prop:finiter} does not apply directly, since a primary learner's posterior expectation is at least as accurate as her prior and is not a location--scale function of the state. It may also be	that primary learners' posteriors are accurate enough regardless of their sampling choices that confirmatory sampling is never optimal for secondary learners.} It does suggest that censoring by primary learners can make secondary learners more inclined to censor. If the group of primary learners with precise priors switched from uncensored to censored sampling, its posteriors would become more accurate while those of the diffuse-prior group would not, widening the gap in accuracy between the two groups' posteriors that confirmatory sampling leverages. The analogous change in the main model is a more informative high type, holding the prior, the low type, and $h$ fixed. The proof of Proposition \ref{prop:finiter}.\ref{prop:finitera} shows that as low types become uninformative, the gain guaranteed by censoring increases with the value of high-type information alone. Selective censoring by primary learners can thus increase heterogeneity in the quality of their posterior beliefs. This should  allay concerns that agents would never use confirmatory sampling if they knew their sources were learning from echo chambers of their own. Mistrust of differing views can be reasonable, and it can also be contagious.

\section{Conclusion}\label{sec:conclusion}

This paper studies conditions under which confirmation bias is rational. When some people's beliefs are more accurate than others, it is reasonable to think those who agree with you are more likely to be among the well-informed (Proposition \ref{prop:hwithinr}). So if time or attention is dear, it may be better to focus attention on views similar to one's own (Proposition \ref{prop:finiter}). As such, exposing people to contrary viewpoints may actually limit their willingness to change their own, if they have little reason to trust what they are hearing. 

Importantly, this does not mean that confirmation bias indicates a healthy environment for learning from others. On the contrary, it signals the presence of these two key frictions: people do not have enough time for all the information out there, and are not sure whom to trust. The selective exposure induced by these constraints may lead people to miss the information that would be most useful to them (Corollary \ref{cor:ineq}). 

So what interventions can help? If high-quality information sources can be identified (marking the social media accounts of reporters from trusted news organizations, for example), then this would obviate the need for filtering based on signal content. Decreasing the proportion of low-quality sources (e.g. those spreading misinformation) would improve inference, though it would not preclude confirmation bias---even if almost all sources are high quality, focusing attention on less extreme signals is preferable if the low-type signal variance is high enough (Proposition \ref{prop:finiter}). Finally, improving the quality of low-type sources would reduce the incentive to filter them out. And easing the agent's time or attention constraint would also reduce the need for filtering. However, these may be harder to translate into concrete policy recommendations.

Our work demonstrates that demand for information can differ depending on its content, not because of preferences but solely to facilitate inference. An important question not addressed here is how the supply side reacts to this phenomenon in equilibrium. Since signals closer to one's prior expectation can elicit stronger reactions than extreme signals (Proposition \ref{prop:actionnonmon}), those seeking to persuade others with distant beliefs may find compromise more effective than accurately stating their views. However, it is not clear that this would persist in equilibrium. Further exploration of the supply side could also extend the binary state in \citet{gentzkow2006media} to a continuum to explore how consumer beliefs shape the range of views that news producers\footnote{These may include news aggregators capable of tailoring their product based on consumer characteristics---see \citet{hu2023politics}.} see fit to print. Understanding whether welfare is improved by news curation (our finding) versus decreased by pandering (theirs) could be a high priority for both theory and empirical work.

Another interesting direction might be distributional concerns in a dynamic setting. While selective exposure can improve beliefs on average, it does make those with inaccurate priors more likely to encounter inaccurate information (Corollary \ref{cor:ineq}). If the radically misinformed are unable to escape their echo chambers and suffer outsize consequences or take actions that negatively affect others, this may outweigh any benefits of accuracy that selective exposure has for others.

Our results yield a novel, testable implication: restricting someone's ability to sample information based on content can reduce the accuracy of their beliefs. Understanding the constraints that lead people to focus their attention on confirmatory views can also shed new light on existing empirical findings and provide important guidance for future work. When those in an echo chamber have less accurate beliefs than those with a more diverse diet of news (as found by e.g. \cite{cookson2020echo}), one interpretation is that the echo chamber is a costly mistake. But is this difference in outcomes driven by differences in cognitive bias, or differences in constraints? Our results show that some people might favor confirmatory news because of limits on time or attention. If so, exposing them to more diverse viewpoints may actually hamper their learning. Distinguishing cognitive bias from constraints may require randomization of information sources---otherwise, different choices may simply reflect different constraints.\footnote{\citet{bail2018exposure} provide an example of such randomization.} Which mechanism dominates is ultimately an empirical question, as is the degree to which beliefs are improved by selective exposure. Future work might seek to measure these forces in various contexts.

\section*{Appendix}
\renewcommand{\thesection}{A\arabic{section}}
\setcounter{section}{0}

\section{Proofs and results not in the main text}\label{app:proofs}

\citet{bail2018exposure} find that disagreement between partisans on Twitter (currently called X) increased more for a group randomly treated with incentives to follow politically opposed accounts. Using their data (\cite{DVN/NSESEH_2018}), Table \ref{tab:bail2018} shows mean ideology (on a seven-point scale, where higher is more conservative) by initial party affiliation and treatment across the course of the study. Disagreement within the control group increased during the study (from 3.16 to 3.21), indicating the presence of forces other than treatment affecting disagreement between the parties. To be clear, our model does not generally predict increasing disagreement over time. However, it can explain why control group disagreement increased just half as much as treatment group disagreement. These results are consistent with trusted (control group) Twitter exposure having a liberal influence on the beliefs of both Democrats and Republicans, the latter of whom would otherwise have shifted to the right under the influence of other (non-Twitter) sources of news and opinion. By reducing trust in views encountered on Twitter, the treatment resulted in a greater increase in disagreement compared to the control group.
\begin{table}[H]
\caption{Mean ideology index over time (\cite{bail2018exposure})}\label{tab:bail2018}
\centering	\begin{tabular}{llll}
		&             & Wave 1 & Wave 5 \\ \hline
		& Republicans & 5.45   & 5.45   \\
		Control & Democrats   & 2.29   & 2.25   \\ \cline{2-4} 
		& Disagreement (R-D)  & 3.16   & 3.21   \\ \hline
		& Republicans   & 5.45   & 5.55   \\
		Treated & Democrats  & 2.36   & 2.35   \\ \cline{2-4} 
		& Disagreement (R-D)  & 3.09   & 3.20   \\ \hline
	\end{tabular}
	\medskip

	\footnotesize{\emph{Note: Ideology measured a seven-point scale where higher is more conservative. Data from \cite{DVN/NSESEH_2018}.}}
\end{table}

The remainder of this appendix provides further detail and results for our model, discussed in the main text. We now proceed with proving the results found in the main text, organized in the order in which their associated results appear in the paper. 

\begin{proof}[Proof of Lemma \ref{lem:brevis} (in Section \ref{sec:results})]
	From the law of total variance,
	\begin{equation}
		\Var\left(\omega\right) = \mathbb{E}\left[\Var\left(\omega\vert r,s\right)\right] + \Var\left(\mathbb{E}\left[\omega\vert r,s\right]\right)
	\end{equation}
	By the definitions of $\sigma_0^2$ and $EU\left(r\right)$ and the fact that the optimal action is the expected state, and substituting each term:
	\begin{equation}
		\sigma_0^2 = -EU\left(r\right) + \Var\left(a^\star\left(s, r\right)\right).
	\end{equation}
	Applying the law of iterated expectation, $\mathbb{E}\left[ a^\star\left(s, r\right)\right]=\omega_0$. Then rearranging yields the result:
	\begin{equation}
		EU\left(r\right) = -\sigma_0^2 + \mathbb{E}\left[\left(a^\star\left(s, r\right)-\omega_0\right)^2\right].
	\end{equation}
\end{proof}
\begin{proof}[Proof of Lemma \ref{lem:weighted_a} (in Section \ref{sec:results})]
		Fix $r$ and a signal realization $s\in\left[\omega_0-r,\omega_0+r\right]$ (realizations outside the sampling radius occur with probability zero). Using $P_r(\omega)\equiv F(\omega_0+r\vert\omega)-F(\omega_0-r\vert\omega)$ for the denominator of Eq.~\ref{eq:truncatedSignal}, we have that $T_rf(s\vert\omega)=f(s\vert\omega)/P_r(\omega)$ in the sampling radius. Since $f=hf_H+(1-h)f_L$, Eq.~\ref{eq:posterior} gives
	\equann
	\mu_1(\omega\vert s,r)&=&\frac{\left[hf_H(s\vert\omega)+(1-h)f_L(s\vert\omega)\right]\mu_0(\omega)/P_r(\omega)}{\int \left[hf_H(s\vert\omega^\prime)+(1-h)f_L(s\vert\omega^\prime)\right]\mu_0(\omega^\prime)/P_r(\omega^\prime)\,d\omega^\prime}\\
	&=&P(H\vert s,r)\,\mu_1(\omega\vert s,r,H)+P(L\vert s,r)\,\mu_1(\omega\vert s,r,L).
	\fequann
	The second line splits the numerator by source type and multiplies and divides the type-$q$ term by $\int f_q(s\vert\omega^\prime)\mu_0(\omega^\prime)/P_r(\omega^\prime)\,d\omega^\prime$. The resulting components are exactly the posteriors conditional on type $\mu_1(\omega\vert s,r,q)$ defined below Eq.~\ref{eq:optimalagiventype}. The resulting weights,
	$$P(H\vert s,r)=\frac{h\int f_H(s\vert\omega^\prime)\mu_0(\omega^\prime)/P_r(\omega^\prime)\,d\omega^\prime}{\int f(s\vert\omega^\prime)\mu_0(\omega^\prime)/P_r(\omega^\prime)\,d\omega^\prime}$$
	and $P(L\vert s,r)=1-P(H\vert s,r)$, are the posterior probabilities of the report's type. That is, the second line is the law of total probability over the type of the observed report, with weights and components computed per Eq. \ref{eq:truncatedSignal}. Now we can multiply both sides by $\omega$ and integrate; by the linearity of the integral and Eq.~\ref{eq:optimalagiventype}, we have 
	$$a^\star(s,r)=P(H\vert s,r)\,a^\star_H(s,r)+P(L\vert s,r)\,a^\star_L(s,r).$$
\end{proof}

\begin{proof}[Proof of Lemma \ref{lem:Hbetter} (in Section \ref{sec:results})]
	The expected utility of an agent who knows the type of the signal she receives is $hEU_H\left(r=\infty\right) + \left(1-h\right)EU_L\left(r=\infty\right) $, so the expected utility of an agent who does not know the type can be no more than this.
\end{proof}

\begin{proof}[Proof of Proposition \ref{prop:hwithinr} (in Section \ref{sec:results})]
	The denominator of Equation \ref{eq:truncatedSignal} is  $hP_H\left(\omega,r\right) +\left(1-h\right)P_L\left(\omega,r\right)$. Given a state $\omega$, the agent draws one report from this sub-population, which is high-type with probability $hP_H\left(\omega,r\right)/\left(hP_H\left(\omega,r\right)+\left(1-h\right)P_L\left(\omega,r\right)\right)$. The chance of a low-type report falling within the sampling radius can be bounded: $P_L\left(\omega,r\right)\leq2r\sup_{\left[\omega_0-r,\omega_0+r\right]}f_L\left(s\vert\omega\right)$. By Equation \ref{eq:fzero}, the right-hand side goes to 0 for every $\omega$ as $\sigma_L^2$ grows large, while $P_H\left(\omega,r\right)>0$ does not vary with $\sigma_L^2$. Then, by dominated convergence (the integrand is bounded by one),
$$
\lim_{\sigma_L^2 \to \infty} \int \mu_0\left(\omega\right) \frac{hP_H\left(\omega,r\right)}{hP_H\left(\omega,r\right)+\left(1-h\right)P_L\left(\omega,r\right)} d\omega \nonumber\\
	= \int \mu_0\left(\omega\right) \frac{hP_H\left(\omega,r\right)}{hP_H\left(\omega,r\right)} d\omega = 1.$$
	
	The same limit holds for the chance that an uncensored report falling within the sampling radius is high-type, $\int\mu_0 hP_H\,d\omega\big/\int\mu_0\left(hP_H+\left(1-h\right)P_L\right)d\omega$. By dominated convergence, the low-type term in the denominator vanishes, while the high-type term is positive and does not vary with $\sigma_L^2$. At finite $\sigma_L^2$ the two probabilities differ, since the latter weights each state by its chance of producing a report within the radius.
\end{proof}

\begin{proof}[Proof of Proposition \ref{prop:finiter}.\ref{prop:finitera} (in Section \ref{sec:results})]
	We will show that if $\sigma_L$ is high enough, there exists a feasible strategy involving censoring which, while not optimal, outperforms the best expected utility achievable without censoring. Since the state $\omega$ has a finite variance $\sigma_0^2$, for any $\epsilon>0$ there exists $r^\prime>0$ such that
	\begin{equation}\label{eq:wgreaterthanr0}
	\sigma_0^2 - \int_{\vert\omega-\omega_0\vert\leq r^\prime}\mu_0\left(\omega\right)(\omega-\omega_0)^2d\omega < \frac{\epsilon}{4}. 
	\end{equation}
	As $\sigma_0^2 = \int_{\vert\omega -\omega_0\vert\leq r^\prime}\mu_0\left(\omega\right)(\omega-\omega_0)^2d\omega + \int_{\vert\omega-\omega_0\vert>r^\prime}\mu_0\left(\omega\right)(\omega-\omega_0)^2d\omega$, this implies 
	\begin{equation}
	\int_{\vert\omega-\omega_0\vert>r^\prime}\mu_0\left(\omega\right)(\omega-\omega_0)^2d\omega < \frac{\epsilon}{4}. \nonumber
	\end{equation}
	Since $r^\prime>0$ implies that $\int_{\vert s-\omega_0\vert>r^\prime} f_H\left(s\vert\omega\right)ds<1$ for any $\omega$,
	\begin{equation}\label{eq:wgreaterthanr}
	\int_{\vert\omega-\omega_0\vert>r^\prime}\int_{\vert s-\omega_0\vert>r^\prime} f_H\left(s\vert\omega\right) \mu_0\left(\omega\right)\left(\omega-\omega_0\right)^2dsd\omega < \frac{\epsilon}{4}. 
	\end{equation}	
	Now consider the signal $s$. The probability distribution of $s$ given a high-type source is the product of $\mu_0\left(\omega\right)$ and $f_H\left(s\vert\omega\right)$ with mean $\omega_0$. Since both $\mu_0$ and $f_H$ have finite variance, by the law of total probability $s$ does as well---denote it $\sigma_{sH}^2$. Then for any $\epsilon>0$ there exists $r^{\prime\prime}$ such that
	\begin{equation}\label{eq:wlessthanr0}
	\sigma_{sH}^2-\int_{-\infty}^\infty\int_{\vert s-\omega_0\vert\leq r^{\prime\prime}} \mu_0\left(\omega\right)f_H\left(s\vert\omega\right) (s-\omega_0)^2dsd\omega < \frac{\epsilon}{4}. 
	\end{equation} 
	So
	\begin{equation}
	\int_{-\infty}^\infty\int_{\vert s-\omega_0\vert>r^{\prime\prime}} \mu_0\left(\omega\right)f_H\left(s\vert\omega\right) \left(s-\omega_0\right)^2dsd\omega < \frac{\epsilon}{4}. \nonumber
	\end{equation}
	Since all terms inside the integral are positive, constricting the domain of integration can only reduce it, and we have
	\begin{equation}
	\int_{\vert\omega-\omega_0\vert\leq r^{\prime\prime}}\int_{\vert s-\omega_0\vert>r^{\prime\prime}} \mu_0\left(\omega\right)f_H\left(s\vert\omega\right) (s-\omega_0)^2dsd\omega < \frac{\epsilon}{4}. \nonumber
	\end{equation}
	Over the domain of integration $(\omega-\omega_0)^2<(s-\omega_0)^2$, so this implies that 
	\begin{equation}\label{eq:wlessthanr}
	\int_{\vert\omega-\omega_0\vert\leq r^{\prime\prime}}\int_{\vert s-\omega_0\vert>r^{\prime\prime}} \mu_0\left(\omega\right)f_H\left(s\vert\omega\right) (\omega-\omega_0)^2dsd\omega < \frac{\epsilon}{4}. 
	\end{equation}
	Letting $k$ be the maximum of $r^\prime$ and $r^{\prime\prime}$ and summing inequalities \ref{eq:wgreaterthanr} and \ref{eq:wlessthanr} (with $r^\prime, r^{\prime\prime}$ replaced by $k$), we have
\begin{equation}\label{eq:outofk}
   \int_{-\infty}^\infty\int_{\vert s-\omega_0\vert>k}
	   \mu_0\left(\omega\right)f_H\left(s\vert\omega\right)
	   \left(\omega-\omega_0\right)^2 ds\,d\omega < \frac{\epsilon}{2}.
\end{equation}

	Define the following decision rule, which is feasible for the agent after censoring to any radius $r\geq k$:
\begin{equation}\label{eq:clippedrule}
	\tilde{a}\left(s\right)\equiv\left\{
	\begin{array}{c l}
		a_H^\star\left(s,\infty\right) &; \vert s-\omega_0\vert\leq k\\
		\omega_0 &; \vert s-\omega_0\vert> k.
	\end{array}
	\right.
\end{equation}
Since $f_H$ is bounded and continuous in $s$, $\vert a_H^\star\left(s,\infty\right)\vert$ is bounded on the compact interval $\left[\omega_0-k,\omega_0+k\right]$, so $\overline{a}\equiv\max\left\{\sup_{\vert s-\omega_0\vert\leq k}\vert a_H^\star\left(s,\infty\right)\vert,\vert\omega_0\vert\right\}<\infty$ and
\begin{equation}\label{eq:envelope}
	\vert u\left(\tilde{a}\left(s\right),\omega\right)\vert
	=\left(\tilde{a}\left(s\right)-\omega\right)^2
	\leq 2\overline{a}^2+2\omega^2
	\qquad\text{for all }s,\omega.
\end{equation}
The rule $\tilde{a}$ does not depend on $r$ or $\sigma_L^2$. Its expected utility without censoring and with high types only is
\begin{equation}\label{eq:clippeduncensored}
	\int\int \mu_0\left(\omega\right)f_H\left(s\vert\omega\right)u\left(\tilde{a}\left(s\right),\omega\right)ds\,d\omega
	\;\geq\; V_H-\frac{\epsilon}{2},
\end{equation}
where $V_H\equiv EU_H\left(r=\infty\right)=\int\int\mu_0\left(\omega\right)f_H\left(s\vert\omega\right)u\left(a_H^\star\left(s,\infty\right),\omega\right)ds\,d\omega$. To see this, split the left side at $\vert s-\omega_0\vert=k$. On the inner region $\tilde{a}=a_H^\star\left(\cdot,\infty\right)$, and since $u\leq0$ the inner integral is no smaller than the same integral taken over all of $\mathbb{R}$, which is $V_H$. On the outer region $\tilde{a}=\omega_0$, and the contribution is $-\int\int_{\vert s-\omega_0\vert>k}\mu_0 f_H\left(\omega-\omega_0\right)^2ds\,d\omega>-\epsilon/2$ by Eq. \ref{eq:outofk}.

Now consider censoring. For $q\in\{H,L\}$ and $r\geq k$, define the type-$q$ sampling region mass (also used in Prop. \ref{prop:hwithinr}) and utility contribution given the state:
\begin{equation*}
	P_q\left(\omega,r\right) \equiv \int_{\vert s-\omega_0\vert\leq r}f_q\left(s\vert\omega\right)ds,
	\qquad
	U_q\left(\omega,r\right) \equiv \int_{\vert s-\omega_0\vert\leq r}f_q\left(s\vert\omega\right)u\left(\tilde{a}\left(s\right),\omega\right)ds.
\end{equation*}
Note that $U_L$ and $P_L$ depend on $\sigma_L^2$ while $U_H$ and $P_H$ do not, and that $P_H\left(\omega,r\right)>0$ for any $\omega$, since $f_H>0$. By Eq. \ref{eq:truncatedSignal}, the density of the sampled signal given the state is $f\left(s\vert\omega\right)$ divided by the sampling region mass $F\left(\omega_0+r\vert\omega\right)-F\left(\omega_0-r\vert\omega\right)=hP_H\left(\omega,r\right)+\left(1-h\right)P_L\left(\omega,r\right)$, so by Eqs. \ref{eq:V}--\ref{eq:objective} the expected utility of censoring to $r$ and playing $\tilde{a}$ with both types is
\begin{equation}\label{eq:statewiseEU}
	EU_{\tilde{a}}\left(r\right)
	=\int\mu_0\left(\omega\right)\,
	\frac{hU_H\left(\omega,r\right)+\left(1-h\right)U_L\left(\omega,r\right)}
	{hP_H\left(\omega,r\right)+\left(1-h\right)P_L\left(\omega,r\right)}\,d\omega,
\end{equation}
with the sampling region masses inside the state integral; with high types only it would be $\int\mu_0\left(\omega\right)\frac{U_H\left(\omega,r\right)}{P_H\left(\omega,r\right)}d\omega$. Since for each $\omega$ the ratio in Eq. \ref{eq:statewiseEU} is an average of $u\left(\tilde{a}\left(s\right),\omega\right)$ against a density on the sampling region, Eq. \ref{eq:envelope} gives
\begin{equation}\label{eq:ratiobound}
	\left\vert\,\frac{hU_H+\left(1-h\right)U_L}{hP_H+\left(1-h\right)P_L}\,\right\vert
	\leq 2\overline{a}^2+2\omega^2,
	\qquad
	\left\vert\,\frac{U_H}{P_H}\,\right\vert
	\leq 2\overline{a}^2+2\omega^2,
\end{equation}
uniformly in $r$ and $\sigma_L^2$, and $2\overline{a}^2+2\omega^2$ is integrable with respect to $\mu_0$ since $\sigma_0^2<\infty$.

To address the choice of $r$, consider a fixed $\omega$. As $r\rightarrow\infty$, $P_H\left(\omega,r\right)\rightarrow1$ and, by dominated convergence in $s$ (with dominating function $\left(2\overline{a}^2+2\omega^2\right)f_H\left(s\vert\omega\right)$), $U_H\left(\omega,r\right)\rightarrow\int f_H\left(s\vert\omega\right)u\left(\tilde{a}\left(s\right),\omega\right)ds$. So the high-types-only integrand converges pointwise to $$\int f_H\left(s\vert\omega\right)u\left(\tilde{a}\left(s\right),\omega\right)ds.$$ By Eq. \ref{eq:ratiobound} and dominated convergence in $\omega$,
\begin{equation*}
	\lim_{r\rightarrow\infty}\int\mu_0\left(\omega\right)\frac{U_H\left(\omega,r\right)}{P_H\left(\omega,r\right)}d\omega
	=\int\int\mu_0\left(\omega\right)f_H\left(s\vert\omega\right)u\left(\tilde{a}\left(s\right),\omega\right)ds\,d\omega
	\;\geq\; V_H-\frac{\epsilon}{2}
\end{equation*}
by Eq. \ref{eq:clippeduncensored}. Fix $r\geq k$ large enough that
\begin{equation}\label{eq:pickr}
	\int\mu_0\left(\omega\right)\frac{U_H\left(\omega,r\right)}{P_H\left(\omega,r\right)}d\omega
	\;\geq\; V_H-\frac{3\epsilon}{4}.
\end{equation}

Finally, we are ready to take the limit in $\sigma_L^2$. Fix $\omega$ (and the $r$ just chosen). By Eq. \ref{eq:fzero},
\begin{equation*}
	P_L\left(\omega,r\right)\leq 2r\sup_{\vert s-\omega_0\vert\leq r}f_L\left(s\vert\omega\right)\rightarrow0,
	\qquad
	\vert U_L\left(\omega,r\right)\vert\leq\left(2\overline{a}^2+2\omega^2\right)P_L\left(\omega,r\right)\rightarrow0
\end{equation*}
as $\sigma_L^2\rightarrow\infty$. So the integrand of Eq. \ref{eq:statewiseEU} converges pointwise to $\frac{hU_H\left(\omega,r\right)}{hP_H\left(\omega,r\right)}=\frac{U_H\left(\omega,r\right)}{P_H\left(\omega,r\right)}$ (the proportion $h$ cancels state by state). By Eq. \ref{eq:ratiobound} and dominated convergence (applied along an arbitrary sequence $\sigma_L^2\rightarrow\infty$),
\begin{equation*}
	EU_{\tilde{a}}\left(r\right)\rightarrow\int\mu_0\left(\omega\right)\frac{U_H\left(\omega,r\right)}{P_H\left(\omega,r\right)}d\omega,
\end{equation*}
so by Eq. \ref{eq:pickr} there exists $\nu_1$ such that $EU_{\tilde{a}}\left(r\right)\geq V_H-\epsilon$ for all $\sigma_L^2\geq\nu_1$. Since $\tilde{a}$ is feasible (though not necessarily optimal) after censoring to $r$,
\[
EU\left(r\right)\geq EU_{\tilde{a}}\left(r\right)\geq V_H-\epsilon
\qquad\text{for all }\sigma_L^2\geq\nu_1.
\]

	So as $r$ and $\sigma_L^2$ grow, the expected utility of censoring to $r$ and playing $a_H^\star\left(\cdot,\infty\right)$ approaches $V_H\equiv EU_H\left(r=\infty\right)$, the expected utility of not censoring with high types only. To make the threshold explicit, define $\Delta\equiv\left(1-h\right)\left(V_H+\sigma_0^2\right)$, which is strictly positive because a high-type signal is valuable: $V_H>-\sigma_0^2$.\footnote{By the identity of Lemma \ref{lem:brevis} applied with high types only, $V_H=-\sigma_0^2+E\left[\left(a_H^\star\left(s,\infty\right)-\omega_0\right)^2\right]$, so allowing $V_H=-\sigma_0^2$ would require $a_H^\star\left(s,\infty\right) = E\left[\omega\vert s\right]=\omega_0$ for almost every $s$, and hence $\mathrm{Cov}\left(s,\omega\right)=0$. But $f_H$ is unbiased, so $\mathrm{Cov}\left(s,\omega\right) =E\left[\omega\,E\left[s\vert\omega\right]\right]-\omega_0^2 =E\left[\omega^2\right]-\omega_0^2=\sigma_0^2>0$---a contradiction.} We can apply the preceding argument with $\epsilon\equiv\Delta/3$: fix $k$ as in Eq. \ref{eq:outofk} and then $r$ and $\nu_1$ as above, so that $EU\left(r\right)\ge V_H-\epsilon$ for all $\sigma_L^2\ge\nu_1$.
	
	What about the value of uncensored sampling? Since by assumption the informativeness of low-type signals goes to zero as their variance grows, $EU_L\left(r=\infty\right)\to-\sigma_0^2$ as $\sigma_L^2\to\infty$, so there exists $\nu_2$ such that $EU_L\left(r=\infty\right)\le-\sigma_0^2+\epsilon$ for all $\sigma_L^2\ge\nu_2$. By Lemma \ref{lem:Hbetter},
	\[
	EU\left(r=\infty\right)\le hV_H+\left(1-h\right)\left(-\sigma_0^2+\epsilon\right)
	\qquad\text{for all }\sigma_L^2\ge\nu_2.
	\]
	Let $\nu^\star\equiv\max\left\{\nu_1,\nu_2\right\}$, enlarged if necessary so that $\nu^\star>\sigma_H^2$. Then for every $\sigma_L^2\ge\nu^\star$,
	\[
	EU\left(r\right)-EU\left(r=\infty\right)
	\ge\left(V_H-\epsilon\right)-hV_H-\left(1-h\right)\left(-\sigma_0^2+\epsilon\right)
	=\Delta-\left(2-h\right)\epsilon
	\ge\Delta-2\epsilon=\frac{\Delta}{3}>0.
	\]
	So for every $\sigma_L^2\ge\nu^\star$, censoring to $r$ yields strictly	higher expected utility than not censoring; in particular, $r=\infty$ is not optimal.
	
\end{proof}

To prove Proposition \ref{prop:finiter}.\ref{prop:finiterb} we will need the following lemma, which shows that if the distributions associated with the agent's problem are scaled by a factor $c$, the optimal actions will scale accordingly.
\begin{lemma}[Scaling]\label{lem:scaling}
	Fix the agent's problem $P=\left(\mu_0,f_H,f_L,h\right)$ with prior mean $\omega_0=0$,
	and fix $c>0$. Consider the ``rescaled'' problem $P^c=\left(\mu_0^c,f_H^c,f_L^c,h\right)$ where $\mu_0^c\left(\omega\right)\equiv c\,\mu_0\left(c\omega\right)$, and $f_q^c\left(s\vert\omega\right)\equiv c\,f_q\left(cs\vert c\omega\right)$ for $q\in\left\{H,L\right\}$. The actions $\{r,a\left(\cdot\right)\}$ are optimal on the equilibrium path for $P$ iff $\{\frac{r}{c},\frac{a\left(cs\right)}{c}\}$ are optimal for $P^c$.
\end{lemma}

\begin{proof}
	First, the cdf of $f^c\left(\cdot\vert\omega\right)$ is $F^c\left(x\vert\omega\right)=\int_{-\infty}^{x}c\,f\left(ct\vert c\omega\right)dt=F\left(cx\vert c\omega\right)$, so the normalizing probability in $P^c$ at radius $r/c$ and state $\omega$ equals the normalizing probability in $P$ at radius $r$ and state $c\omega$. Substituting into Eq. 2 yields $T_{r/c}f^c\left(s\vert\omega\right)=c\left(T_rf\right)\left(cs\vert c\omega\right)$, where $f\equiv hf_H+\left(1-h\right)f_L$; since rescaling is linear in densities, this holds type by type and for the mixture $f$. In particular, symmetric windows about the prior mean map to symmetric windows about the rescaled prior mean, with $r=\infty$ mapping to $r=\infty$.
	
	Second, the map $\left(r,a\left(\cdot\right)\right)\mapsto\left(r/c,\,a\left(c\,\cdot\right)/c\right)$ is a bijection from the strategies of $P$ onto the strategies of $P^c$: it carries an action rule on $\left(-r,r\right)$ to an action rule on $\left(-r/c,r/c\right)$, and its inverse is the corresponding rescaling by $1/c$.
	
	Third, for any strategy $\left(r,a\left(\cdot\right)\right)$ of $P$, substitute $\omega=c\tilde\omega$ and $s=c\tilde s$, so that $d\tilde s\,d\tilde\omega=c^{-2}\,ds\,d\omega$:
	\begin{align*}
		U_{P^c}\left(r/c,\,a\left(c\tilde s\right)/c\right)
		&=-\iint_{\vert\tilde s\vert\le r/c}\left(a\left(c\tilde s\right)/c-\tilde\omega\right)^2 T_{r/c}f^c\left(\tilde s\vert\tilde\omega\right)\mu_0^c\left(\tilde\omega\right)d\tilde s\,d\tilde\omega\\
		&=-\iint_{\vert s\vert\le r}\frac{\left(a\left(s\right)-\omega\right)^2}{c^2}\,c\left(T_rf\right)\left(s\vert\omega\right)\,c\,\mu_0\left(\omega\right)\frac{ds\,d\omega}{c^2}=c^{-2}\,U_P\left(r,a\right).
	\end{align*}
	So the rescaling multiplies the expected utility of every strategy by the positive constant $c^{-2}$ and places the strategy spaces of $P$ and $P^c$ in one-to-one correspondence. Hence $\left(r,a\left(\cdot\right)\right)$ attains the supremum of $U_P$ if and only if $\left(r/c,\,a\left(c\,\cdot\right)/c\right)$ attains the supremum of $U_{P^c}$, and the optimal radii of $P^c$ are exactly the optimal radii of $P$ divided by $c$; in particular, a finite optimal radius corresponds to a finite optimal radius.
	
\end{proof}

With this lemma in hand, we show that for any location-scale low-type noise $f_L$, if the prior and high-type signals are precise enough, censoring to a finite radius is optimal. 

\begin{proof}[Proof of Proposition \ref{prop:finiter}.\ref{prop:finiterb} (in Section \ref{sec:results})]
	For $\nu\geq 1$ define $g_\nu\left(s\vert\omega\right) \equiv \frac{1}{\nu}f_L\big(\frac{s}{\nu}\vert \frac{\omega}{\nu}\big)$, the member of the family of low-type distributions with variance $\nu^2\sigma_L^2$. Since $f_L$ is continuous, bounded, and strictly positive, $g_\nu$ inherits those properties, and rewriting the supremum of $g$ gives us
	\begin{equation}
		\sup_{s,\omega}g_\nu\left(s\vert\omega\right)
		= \frac{1}{\nu}\sup_{s,\omega}f_L\left(s\vert\omega\right)
		= \frac{\Vert f_L\Vert_\infty}{\nu}\to0
		\quad\text{as }\nu\to\infty,
	\end{equation}
	since $\Vert f_L\Vert_\infty<\infty$ by assumption. So the family $\left\{g_\nu\right\}_{\nu\ge1}$ also satisfies convergence on every bounded interval. We can therefore apply Proposition \ref{prop:finiter}.\ref{prop:finitera} to the problem $\left(\mu_0,f_H,g_\nu,h\right)$: letting
	$\nu^{\star\star}\equiv\max\left\{1,\sqrt{\nu^\star}/\sigma_L\right\}$, we have that $\nu^2\sigma_L^2\ge\nu^\star$ for every $\nu\ge\nu^{\star\star}$, so censoring is optimal. Applying Lemma	\ref{lem:scaling} with $c=\nu$, scaling the problem $\left(\mu_0,f_H,g_\nu,h\right)$ by $\nu$ is exactly $\left(\tilde\mu_0^\nu,\tilde f_H^\nu,f_L,h\right)$, since
	$\nu\,g_\nu\left(\nu s\vert\nu\omega\right)=f_L\left(s\vert\omega\right)$. By the lemma, censoring is optimal for the rescaled problem: in particular, its optimal radius is $r_\nu^\star/\nu<\infty$ if $r^\star$ is optimal for the original problem $\left(\mu_0,f_H,g_\nu,h\right)$.
\end{proof}

We next show that the main results of Proposition \ref{prop:finiter} are robust to a ``leaky'' echo chamber, which occasionally provides signals from outside the chosen sampling radius.

\begin{corollary}\label{cor:leaky}
	Suppose that when the agent chooses $r<\infty$, with chance $1-\epsilon$ the signal realization is drawn from $T_{r}f\left(s\vert\omega\right)$ as before, but with a small chance $\epsilon$ the agent receives a signal realization from outside the sampling radius (distributed $\frac{f\left(s\vert\omega\right)\mathbf{1}_{\vert s-\omega_0\vert>r}}{\int_{\vert s-\omega_0\vert>r}f\left(s^\prime\vert\omega\right)ds}$). Assume that radius $r^\prime$ provides higher expected utility than selecting $r=\infty$ when the sampling radius is leak-proof ($\epsilon=0$). Then there exists $\bar{\epsilon}>0$ such that for any $\epsilon\in\left(0,\bar{\epsilon}\right)$, censoring is optimal. 
\end{corollary}

\begin{proof}
Without loss, let $\omega_0=0$. We show that the following feasible strategy choosing $r^\prime$ outperforms uncensored sampling for some $\epsilon>0$. If the signal realization is inside the sampling radius ($\vert s\vert\leq\vert r^\prime\vert$), play the optimal action as if $\epsilon=0$, that is $a^\star\left(s\vert r=r^\prime, \epsilon=0\right)$. Call the expected utility in this case $\bar{U}_{r^\prime}$. For signal realizations outside the chosen sampling radius, simply play 0. The expected utility of this strategy is $-\epsilon\sigma_0^2 + \left(1-\epsilon\right) \bar{U}_{r^\prime}.$
The first term is the expected utility of playing 0 multiplied by the chance of a signal realization outside the sampling radius. By continuity, there exists $\bar{\epsilon}$ such that $\epsilon<\bar{\epsilon}$ guarantees this is arbitrarily close to $\bar{U}_{r^\prime}$, which by assumption exceeds the expected utility possible with uncensored sampling. The optimal choice of $r$ may not be $r^\prime$, but we have shown it cannot be $\infty$.
\end{proof}

There are two ways we can interpret this result. One is that even if the agent is unable to perfectly control what information she receives, it is still optimal to try to focus on views she agrees with. The second is that if given a choice of how to allocate her attention, the agent will prefer to pay less attention to views she disagrees with. While we leave exploration of optimal weighting for this and other more complicated strategy spaces to future work, we have shown that our main results are not dependent on the ``leak-proof''  strategy space of the main text.

\begin{proof}[Proof of Lemma \ref{lem:sincreasing} (in Section \ref{sec:normaldist})]
	We provide a more general proof, relying only on $f_H$ being a strictly log-concave location family, $f_H\left(s\vert\omega\right)=g\left(s-\omega\right)$ with $\log g$ strictly concave, so that the likelihood ratio below is increasing in $\omega$ (the formula shown is the normal case).

	\begin{equation*}
		\frac{\mu_1\left(\omega\vert s^\prime,r,H\right)}{\mu_1\left(\omega\vert s,r,H\right)}
		\;\propto\; \frac{f_H\left(s^\prime\vert\omega\right)}{f_H\left(s\vert\omega\right)}
		= \exp\left\{\frac{\left(s^\prime-s\right)\left(2\omega-s-s^\prime\right)}{2\sigma_H^2}\right\}.
	\end{equation*}
	This is strictly increasing in $\omega$: higher states are more likely under $s^\prime$, and lower states are more likely under $s$. The posterior given $s^\prime$ dominates the posterior given $s$ in the likelihood-ratio order, and therefore in first-order stochastic dominance \citep{milgrom1981good} (and strictly, since both posteriors have full support). So from Equation \ref{eq:optimalagiventype}, we have that $a^\star_H\left(s,r\right)<a^\star_H\left(s^\prime,r\right)$.
\end{proof}

\begin{proof}[Proof of Lemma \ref{lem:normalodds} (in Section \ref{sec:normaldist})]

	The odds of a high-type source are $$\frac{\text{P}(H|s,r=\infty)}{\text{P}(L|s,r=\infty)}=\frac{h\int f_H(s|\omega)\mu_0(\omega) d\omega}{(1-h)\int f_L(s|\omega)\mu_0(\omega) d\omega}.$$ Since both the numerator and the denominator are integrals of two Gaussian probability density functions, the product is also Gaussian with a well-known closed form. Namely, 
\begin{multline}\label{lem:normaloddspost}
	\int f_q(s|\omega)\mu_0(\omega)d\omega \\ = \frac{1}{\sqrt{2\pi(\sigma_q^2+\sigma_0^2)}}\exp\left\{-\frac{(s-\omega_0)^2}{2(\sigma_q^2+\sigma_0^2)}\right\}\cdot \int \frac{1}{\sqrt{2\pi\sigma_{1|q}^2}}\exp\left\{-\frac{(\omega-\omega_{1|q})^2}{2\sigma_{1|q}^2}\right\}d\omega,
\end{multline}
	for $ q\in\{H,L\}$, where $\sigma_{1|q}^2$ and $\omega_{1|q}$ are constants equal to the updated mean and variance posteriors of $\omega$ after observing $s$ given quality $q$. That is, 
	$$\sigma_{1|q}^2=\frac{1}{\sigma_q^{-2}+\sigma_0^{-2}},\qquad \omega_{1|q}=\omega_0\frac{\sigma_{1|q}^2}{\sigma_0^2}+s\frac{\sigma_{1|q}^2}{\sigma_q^2}.$$
	The integral in the right hand side of equation \ref{lem:normaloddspost} is the integral of a normal distribution, so it equals $1$. Thus,
	$$\frac{\text{P}(H|s,r=\infty)}{\text{P}(L|s,r=\infty)}=\frac{h}{1-h}\frac{\sqrt{2\pi(\sigma_L^2+\sigma_0^2)}}{\sqrt{2\pi(\sigma_H^2+\sigma_0^2)}}\exp\left\{-\frac{(s-\omega_0)^2}{2(\sigma_H^2+\sigma_0^2)}\right\}\exp\left\{\frac{(s-\omega_0)^2}{2(\sigma_L^2+\sigma_0^2)}\right\}$$
	
	Rearranging terms, we obtain the result:
	$$\frac{\text{P} (H|s,r=\infty)}{\text{P} (L|s,r=\infty)}=\left(\frac{h}{1-h}\right)
	\sqrt{\frac{\sigma_L^2+\sigma_0^2}{\sigma_H^2+\sigma_0^2}}\cdot \text{exp}\left\{-\frac{( s-\omega_0)^2(\sigma_L^2-\sigma_H^2)}{2(\sigma_H^2+\sigma_0^2)(\sigma_L^2+\sigma_0^2)}\right\}$$
	which is decreasing in $(s-\omega_0)^2$, and therefore in $|s-\omega_0|$ as well.
	
\end{proof}

\begin{proof}[Proof of Corollary \ref{cor:creduseful} in Section \ref{sec:normaldist}]
	By Lemma \ref{lem:weighted_a}, $ a^\star(s,r)=P(H|s,r)a^\star_H(s,r)+P(L|s,r)a^\star_L(s,r)$. By Eq. \ref{eq:normalPolicy}, $a_q^\star\left(s, \infty\right) = \frac{\sigma_q^2}{\sigma_q^2+\sigma_0^2}\omega_0+\frac{\sigma_0^2}{\sigma_q^2+\sigma_0^2}s $ for $q\in\{L,H\}$. Plugging in $a_H^\star$ and $a_L^\star$ from Eq. \ref{eq:normalPolicy}, rearranging, and using $P\left(H\vert s,\infty\right)+P\left(L\vert s,\infty\right)=1$ yields Eq. \ref{eq:tradeoff}. The monotonicity of $w\left(s\right)$ follows from Lemma \ref{lem:normalodds}. 
\end{proof}

\begin{proof}[Proof of Proposition \ref{prop:rstarnormal} in Section \ref{sec:normaldist}]
	We will write the echo chamber boundaries $b^{\pm} \equiv \omega_{0} \pm r$, the chance of an uncensored signal within it $P_{r}(\omega) \equiv F(b^{+}|\omega) - F(b^{-}|\omega)$, and the conditional density of the observed signal at the echo chamber boundary (given the state) $\lambda_{r}(\omega) \equiv \dfrac{f(b^{+}|\omega) + f(b^{-}|\omega)}{P_{r}(\omega)}$. Throughout this proof the all-normal specification provides that $a^{*}(s,r)$ is smooth in $(s,r)$, and that all interchanges of differentiation, limits, and integration below are valid. The optimal action $a^{*}$ is a ratio of integrals of Gaussian densities against the normal prior, and the density of a boundary signal $\lambda_{r}(\omega)$ grows at most linearly in the state $|\omega|$ (Mills-ratio bounds), so every integrand below is dominated by a polynomial times the normal prior.\footnote{Proposition \ref{prop:rstarnormal} may not actually require the differentiability of $a^{*}$ in $r$---see \citet{milgrom2002envelope}, Corollary 4, applied to the compact family of optimal policies $\{a^{*}(\cdot, \rho)\}$.} 
	
	The agent's expected value of choosing $r$ and $a^\star$ we can write as $U\left(r,a\right)\equiv -\int \mu_{0}(\omega)\, \int \big(a(s) - \omega\big)^{2}\, T_{r}f(s|\omega)\, ds d\omega$, with $EU\left(r\right)=U\left(r,a^\star\left(\cdot,r\right)\right)$. The total derivative of $U$ is the sum of two parts: $\partial_r U\left(r,a\right)\vert_{a=a^\star\left(\cdot,r\right)}$, and a policy channel
	\begin{equation}\label{eq:policychannel}
		\int\!\!\int 2 \big( a^{*}(s,r) - \omega \big)\, \partial_{r} a^{*}(s,r)\, \mu_{0}(\omega)\, \frac{f(s|\omega)}{P_{r}(\omega)}\, ds\, d\omega.
	\end{equation}
	We can interchange the order of integration of this term, and use that the action policy $a^\star$ is optimal: $a^{*}(s,r) = \E(\omega | s, r)$ pointwise (Eq. \ref{eq:optimala}). Doing so, we see that line \ref{eq:policychannel} has no first-order contribution to the total derivative (the envelope theorem):
	$$\int 2\, \partial_{r} a^{*}(s,r)\, m_{r}(s) \Big[ a^{*}(s,r) - \E(\omega | s, r) \Big]\, ds \;=\; 0.$$
	
	So the total derivative of the agent's expected utility is just $\partial_r U\left(r,a\right)\vert_{a=a^\star\left(\cdot,r\right)} = \partial_r EU\left(r\right)$ (eq. \ref{eq:objective}). Using the Leibniz rule and the fact that $\partial_{r} \frac{1}{P_{r}} = -\lambda_{r}/P_{r}$,
	\begin{multline}\label{eq:EUenvelope}
		\partial_r U\left(r,a\right)\vert_{a=a^\star\left(\cdot,r\right)} \;=\; \\
		-\sum_{b \in \{b^{\pm}\}} \int \mu_{0}(\omega) \big( a(b) - \omega \big)^{2}\, \frac{f(b|\omega)}{P_{r}(\omega)}\, d\omega
		\;+\; \int \mu_{0}(\omega)\, \lambda_{r}(\omega) \int_{b^{-}}^{b^{+}} \big( a(s) - \omega \big)^{2}\, \frac{f(s|\omega)}{P_{r}(\omega)}\, ds\, d\omega.
	\end{multline}
	
	Recalling that $m_{r}(s) \equiv \int \mu_{0}(\omega)\, T_{r} f(s|\omega)\, d\omega$, the posterior expected state given a signal at the echo chamber boundary $b$ is $\mu_{1}(\omega | b, r) = \frac{\mu_{0}(\omega) f(b|\omega)}{  P_{r}(\omega)\, m_{r}(b) }$, and $a^{*}(b,r)$ is its mean. So the summation terms of eq. \ref{eq:EUenvelope} at $b^\pm$  (evaluated at $a = a^{*}(\cdot, r)$) equal $m_{r}(b^\pm)\, \Var(\omega | b\pm, r)$. Turning to the second term of Eq. \ref{eq:EUenvelope},   recognize that the inner integral $ds$ is $V\left(\omega,r\right)$. Writing $\lambda_{r} = \sum_{b \in \{b^{\pm}\}} f(b|\cdot)/P_{r}$ gives $\sum_{b \in \{b^{\pm}\}} m_{r}(b)\,\E \big( V\left(\omega,r\right), \big|\, s = b,\, r \big)$. Putting the two together, Eq. \ref{eq:EUenvelope} becomes
	\begin{equation}
		EU'(r) \;=\; \sum_{b \in \{b^{\pm}\}} m_{r}(b) \Big[\; \E\big( V\left(\omega,r\right) \, \big|\, s = b,\, r \big) \;-\; \Var(\omega | b, r) \;\Big].
	\end{equation}
	
	Since both $\mu_0$ and $f_{q}(\cdot|\omega)$ are normal, reflecting both the state $\omega$ and signal $s$ about $\omega_0$  leaves the density $\mu_0\left(\omega\right)f\left(s\vert\omega\right)/P_r\left(\omega\right)$ of the state-signal pair unchanged: the reflection negates $s-\omega$, and the echo chamber is symmetric about $\omega_0$. So swapping $b^+$ and $b^-$, the two summed terms are equal, giving Eq. \ref{eq:EUprime}.
	
	For any compact $C\subset\R$, as $r\downarrow 0$, the ratios $f\left(s\vert\omega\right) / f\left(\omega_0\vert\omega\right)$ and $P_r\left(\omega\right)/\left[2rf\left(\omega_0\vert\omega\right)\right]$ converge to one uniformly over $\vert s-\omega_0\vert\leq r$ and $\omega\in C$. So $2rf\left(s\vert\omega\right)/P_r\left(\omega\right)$ goes to one as well, and $\mu_1\left(\cdot\vert s,r\right)\rightarrow\mu_0$. Nothing is learned from a report from this vanishing echo chamber: $a^\star \rightarrow \omega_{0}$, and by Lemma \ref{lem:brevis} with dominated convergence, $EU(r) \to -\sigma_{0}^{2}$. As $r \rightarrow \infty$, $P_{r} \to 1$ and $EU(r) \to EU(\infty)$.  Since the integrand of Eq. \ref{eq:EUinfty} is bounded below by $\big[\sigma_{0}^{2}/(\sigma_{0}^{2} + \sigma_{L}^{2})\big]^{2} (s - \omega_{0})^{2}\, m(s)$ (see Eq. \ref{eq:tradeoff}),  $EU(\infty) > -\sigma_{0}^{2}$, and thus $EU(r) > -\sigma_{0}^{2}$ in the limit. Under the conditions of Proposition \ref{prop:finiter} there is a finite $\hat{r}$ with $EU(\hat{r}) > EU(\infty)$. Both endpoint limits therefore lie strictly below $EU(\hat{r})$, so the supremum $r^{\star}$ of the continuous function $EU$ lies in a compact subinterval of $(0, \infty)$, where its first derivative (Eq. \ref{eq:EUprime}) vanishes: $\Var(\omega\,|\,s = b^{+}, r^{\star}) = \E(V(\omega, r^{\star})\,|\,s = b^{+}, r^{\star})$. Subtracting each side from $\E[(\omega - \omega_{0})^{2}\,|\,s = b^{+}, r^{\star}]$ and using the bias--variance identity $\Var(\omega|s,r) = \E[(\omega - \omega_{0})^{2}|s,r] - (a^{*}(s,r) - \omega_{0})^{2}$ yields Eq. \ref{eq:tradeoffatrstar}. 
\end{proof}

\begin{proof}[Proof of Proposition \ref{prop:actionnonmon} (in Section \ref{sec:normaldist})]
	Without loss, fix $\omega_0=0$. Recall that by Lemma \ref{lem:normalodds}, $$\frac{\text{P} (H|s,r=\infty)}{\text{P} (L|s,r=\infty)}=\left(\frac{h}{1-h}\right)
	\sqrt{\frac{\sigma_L^2+\sigma_0^2}{\sigma_H^2+\sigma_0^2}}\cdot \text{\em{exp}}\left\{-\frac{ s^2(\sigma_L^2-\sigma_H^2)}{2(\sigma_H^2+\sigma_0^2)(\sigma_L^2+\sigma_0^2)}\right\}.$$
	By Corollary \ref{cor:creduseful}, we have $
	a^{*}(s, \infty) \;=\; w(s)\, s$, where $w(s) \;\equiv\; P(H|s)\, \frac{\sigma_{0}^{2}}{\sigma_{0}^{2} + \sigma_{H}^{2}} \;+\; P(L|s)\, \frac{\sigma_{0}^{2}}{\sigma_{0}^{2} + \sigma_{L}^{2}}$.
	
	Fix $\sigma_0^2$ and $\sigma_H^2$. Choose a positive constant $C>1$ and fix a signal $s$ such that $s\equiv C\left(\sigma_H^2 + \sigma_0^2\right)$. By Lemma \ref{lem:normalodds}, as $\sigma_L^2\to\infty$ the odds of a high type given signal $s$ go to $\infty$, so by Corollary \ref{cor:creduseful}, $a^\star\left(s,\infty\right) \to \left(s-\omega_0\right)\frac{\sigma_{0}^{2}}{\sigma_{0}^{2} + \sigma_{H}^{2}}$. 
	
	Next, let $s^\prime \equiv \sigma_L^2$. As $\sigma_L^2$ grows, the odds of a high type given $s^\prime$ go to zero by Lemma \ref{lem:normalodds}. So we have that $a^\star\left(s^\prime,\infty\right)$ goes to $s^\prime\frac{\sigma_0^2}{\sigma_0^2 + \sigma_L^2}$, and plugging in $s^\prime$ and taking the limit as $\sigma_L^2\to\infty$ gives us $a^\star\left(s^\prime,\infty\right)\to\sigma_0^2$. So as $\sigma_L^2\to\infty$, we have $a^\star\left(s,\infty\right) - a^\star\left(s^\prime,\infty\right) \to s\frac{\sigma_{0}^{2}}{\sigma_{0}^{2} + \sigma_{H}^{2}} - \sigma_0^2 = \left(C-1\right)\sigma_0^2>0$. Since $s^\prime>s$ as $\sigma_L^2$ grows, this gives us non-monotonicity: there exists $\bar{\sigma_L^2}$ such that for all $\sigma_L^2>\bar{\sigma_L^2}$, $a^\star\left(s,\infty\right) > a^\star\left(s^\prime,\infty\right)$. A local maximum of $a^\star$ over the compact interval $\left[\omega_0,s^\prime\right]$ is attained by some interior signal, since by Corollary \ref{cor:creduseful} $a^\star$ is continuous in $s$ and equal to $0$ at $s=\omega_0$, which (along with $a^\star\left(s^\prime,\infty\right)$) is less than $a^\star\left(s,\infty\right)$.
\end{proof}

To prove Proposition \ref{prop:actionlin} we will need Lemma \ref{lem:onetype}, which shows that if there is just one source type, un-censored sampling is uniquely optimal.

\begin{lemma}\label{lem:onetype}
	Assume $\mu_0$ is normal and that all sources are of a single type that has normal noise $f_q$ of variance $\sigma_q^2$. Then $EU(r)<EU(\infty)$ for every $r<\infty$.
\end{lemma}

\begin{proof}
	Fix $r<\infty$ and write $P_r\left(\omega\right)\equiv F_q\left(\omega_0+r\vert\omega\right)-F_q\left(\omega_0-r\vert\omega\right)$ for the denominator in Equation \ref{eq:truncatedSignal}, so that $T_rf_q\left(s\vert\omega\right)=f_q\left(s\vert\omega\right)/P_r\left(\omega\right)$ for $s\in\left(\omega_0-r,\omega_0+r\right)$. 

	We will use the \citet{gill1995applications} rendition (Eq. 3)\footnote{While stated for a closed interval, the result also applies to $\mathbb{R}$---see \citet{gassiat2024vantrees}.} of \citet{vantrees1968}, which tells us that for any action policy $a\left(s\right)$ (including the optimal policy $a^\star\left(\cdot,r\right)$),
	\begin{equation}\label{eq:vantrees}
		\E\left[\left(a\left(s\right)-\omega\right)^2\right]\;\geq\;\frac{1}{\E\left[\mathcal{I}_r\left(\omega\right)\right]+\mathcal{I}\left(\mu_0\right)},
	\end{equation}
	where the outer expectations are taken over the joint distribution $\mu_0\left(\omega\right)T_rf_q\left(s\vert\omega\right)$ of the state and the censored signal. Here
	$$\mathcal{I}_r\left(\omega\right)\equiv\E\left[\left(\partial_\omega\ln T_rf_q\left(s\vert\omega\right)\right)^2\,\Big\vert\,\omega\right] \qquad\mbox{ and }\qquad \mathcal{I}\left(\mu_0\right)\equiv\E\left[\left(\partial_\omega\ln\mu_0\left(\omega\right)\right)^2\right]=\frac{1}{\sigma_0^2}$$
	are the Fisher information that the censored signal contains about the state and the Fisher information of the prior, respectively. By Equations \ref{eq:V} and \ref{eq:objective}, the left side of Equation \ref{eq:vantrees} evaluated at $a=a^\star\left(\cdot,r\right)$ is $-EU\left(r\right)$.
	
	Applying \citet{gill1995applications} requires that $\mu_0$ and $\omega\mapsto T_rf_q\left(s\vert\omega\right)$ are absolutely continuous and that the score $\partial_\omega\ln T_rf_q\left(s\vert\omega\right)$ has mean zero conditional on state $\omega$. These are properties of the density family and place no restriction on the action policy, so the bias of $a^\star\left(\cdot,r\right)$ is immaterial. Both densities are smooth in $\omega$ (for $T_r f_q$, because the sampling interval's endpoints do not depend on $\omega$), which supplies absolute continuity. The score has mean zero conditional on $\omega$ because
	$$\int_{\omega_0-r}^{\omega_0+r}\partial_\omega T_rf_q\left(s\vert\omega\right)ds=\partial_\omega\int_{\omega_0-r}^{\omega_0+r}T_rf_q\left(s\vert\omega\right)ds=\partial_\omega 1=0,$$
	allowed because the limits of integration do not depend on the state $\omega$. 
	
	We now compute $\mathcal{I}_r\left(\omega\right)$. All expectations below, and $\E\left[s\vert\omega\right]$ in particular, are taken under $T_rf_q\left(\cdot\vert\omega\right)$. The score is $\partial_\omega\ln T_rf_q\left(s\vert\omega\right)=\frac{s-\omega}{\sigma_q^2}-\partial_\omega\ln P_r\left(\omega\right)$. The second term $\partial_\omega\ln P_r\left(\omega\right)$ does not depend on $s$, so taking conditional means of both sides and then using that the score has mean zero conditional on $\omega$ shows that $\partial_\omega\ln P_r\left(\omega\right)=\left(\E\left[ s\vert\omega\right] - \omega\right)/\sigma_q^2$. Canceling $\omega$ yields that the score is  $\left(s-\E\left[s\vert\omega\right]\right)/\sigma_q^2$, and the conditional expectation of its square is
	\begin{equation}\label{eq:fisherCensored}
		\mathcal{I}_r\left(\omega\right)=\frac{\Var\left(s\vert\omega\right)}{\sigma_q^4},
	\end{equation}
	the conditional variance of the censored signal divided by $\sigma_q^4$. We now show that $\Var\left(s\vert\omega\right)<\sigma_q^2$. 
	First, note that $\left(s-\omega\right)\left(s-\E\left[s\vert\omega\right]\right) = \left(s-\E\left[s\vert\omega\right]\right)^2 + \left(\E\left[ s\vert\omega\right] - \omega\right)\left( s - \E\left[ s\vert\omega\right] \right)$. Taking the expectation conditional on $\omega$, the first term becomes $\Var\left(s\vert\omega\right)$ and the second is zero, so  $\Var\left(s\vert\omega\right)=\E\left[\left(s-\omega\right)\left(s-\E\left[s\vert\omega\right]\right)\,\big\vert\,\omega\right]$. Then we have	
	\begin{multline*}
		\Var\left(s\vert\omega\right)=\frac{1}{P_r\left(\omega\right)} \int_{\omega_0-r}^{\omega_0+r}\left(s-\omega\right)\left(s-\E\left[s\vert\omega\right]\right)\mathcal{N}\left(s-\omega;0,\sigma_q^2\right) ds\\
		=\sigma_q^2\left[1-\frac{\left(\omega_0+r-\E\left[s\vert\omega\right]\right)\N\left(\omega_0+r-\omega;0,\sigma_q^2\right)+\left(\E\left[s\vert\omega\right]-\omega_0+r\right)\N\left(\omega_0-r-\omega;0,\sigma_q^2\right)}{P_r\left(\omega\right)}\right],
	\end{multline*}
	where $\mathcal{N}\left(s;\mu,\sigma^2\right)$ is the normal pdf. The second line follows by integration by parts, using $\left(x-\omega\right)\N\left(x-\omega;0,\sigma_q^2\right)=-\sigma_q^2\,\partial_x\N\left(x-\omega;0,\sigma_q^2\right)$, and a little rearranging. Both terms in the numerator are strictly positive because the conditional mean $\E\left[s\vert\omega\right]$ lies strictly inside the sampling interval, so both of the distances from the conditional mean to the endpoints of the interval are positive. So $\Var\left(s\vert\omega\right)<\sigma_q^2$, and by Eq. \ref{eq:fisherCensored} we have $\mathcal{I}_r\left(\omega\right)<1/\sigma_q^2$ for every $\omega$, where $1/\sigma_q^2$ is the Fisher information of an un-censored signal. Therefore, $\E\left[\mathcal{I}_r\left(\omega\right)\right]<1/\sigma_q^2$. Substituting this into Eq. \ref{eq:vantrees}, 
	$$-EU\left(r\right)\;>\;\frac{1}{1/\sigma_q^2+1/\sigma_0^2}\;=\;\frac{\sigma_0^2\sigma_q^2}{\sigma_0^2+\sigma_q^2}.$$
	
	Finally, if $r=\infty$ then Equation \ref{eq:truncatedSignal} is equal to $f_q$ itself, a normal distribution, and standard Bayesian inference yields that the posterior is normal with variance $\sigma_0^2\sigma_q^2/\left(\sigma_0^2+\sigma_q^2\right)$ regardless of $s$, with mean $a^\star\left(s,\infty\right)$ as in Equation \ref{eq:normalPolicy}. So $-EU\left(\infty\right)=\sigma_0^2\sigma_q^2/\left(\sigma_0^2+\sigma_q^2\right)$, which is exactly the bound above, and $EU\left(r\right)<EU\left(\infty\right)$.
\end{proof}

\begin{proof}[Proof of Proposition \ref{prop:actionlin} (in Section \ref{sec:normaldist})]
	We will write $\theta\equiv\left(\sigma_H^2,\sigma_L^2,h\right)$ for the parameters (high- and low-type signal variance, and fraction of high types), and formalize the dependence of expected utility and the optimal action on them by writing $EU\left(r;\theta\right)$ and $a^\star\left(s,r;\theta\right)$. Let $\theta_n\rightarrow\theta^\ast$ be the sequence in the statement. At the limit $\theta^\ast$ there is just a single source type: if $h=1$ or $\sigma_H^2=\sigma_L^2$ the mixture $f=hf_H+\left(1-h\right)f_L$ is just $f_H$, and if $h=0$ it is $f_L$; in either case the surviving type has variance $\sigma_q^2$. By Lemma \ref{lem:onetype}, $r=\infty$ is the unique optimal radius at $\theta^\ast$, and $a^\star\left(s,\infty;\theta^\ast\right)$ is the linear function of $s$ in Equation \ref{eq:normalPolicy} with variance $\sigma_q^2$.
	
	The extension from $\theta^\ast$ to sequences approaching it is a standard continuity argument, limited here to a sketch for brevity. Since the sampling interval is centered at $\omega_0$, the censored signal density of Eq. \ref{eq:truncatedSignal} is jointly continuous in the state, the signal, the radius, and parameters $\theta$ on the interior of the sampling interval. A dominated convergence argument can be used to show that $EU\left(r;\theta\right)$ is jointly continuous in $\left(r,\theta\right)$ on $\bar\R_+\times\left(0,\infty\right)^2\times\left[0,1\right]$, and that $a^\star\left(s,r;\theta\right)$ is jointly continuous in $\left(r,\theta\right)$ for every $s$ inside the sampling interval. Continuity at $r=\infty$ here means (naturally) that $EU\left(r_n;\theta_n\right)\rightarrow EU\left(\infty;\theta\right)$ and $a^\star\left(s,r_n;\theta_n\right)\rightarrow a^\star\left(s,\infty;\theta\right)$ whenever $r_n\rightarrow\infty$ and $\theta_n\rightarrow\theta$. In particular, $EU\left(\cdot;\theta\right)$ is continuous on the compact set $\bar\R_+$, so an optimal radius exists for every set of parameters $\theta$; if the optimal radius is not unique, the argument applies to any selection. Now fix some $M<\infty$. By Lemma \ref{lem:onetype} and continuity in $r$, $\sup_{r\leq M}EU\left(r;\theta^\ast\right)$ is attained and lies strictly below $EU\left(\infty;\theta^\ast\right)$. By joint continuity, $\sup_{r\leq M}EU\left(r;\theta_n\right)\rightarrow\sup_{r\leq M}EU\left(r;\theta^\ast\right)$ and $EU\left(\infty;\theta_n\right)\rightarrow EU\left(\infty;\theta^\ast\right)$, so for all $n$ large enough $EU\left(\infty;\theta_n\right)>\sup_{r\leq M}EU\left(r;\theta_n\right)$ and we have $r^\star_n>M$. Since $M$ was arbitrary, $r^\star_n\rightarrow\infty$. However, note that we do not claim $r^\star_n=\infty$ for large $n$---censoring may remain strictly optimal at every $\theta_n$. Finally, $\left(r^\star_n,\theta_n\right)\rightarrow\left(\infty,\theta^\ast\right)$, so joint continuity of $a^\star$ gives $a^\star\left(s,r^\star_n;\theta_n\right)\rightarrow a^\star\left(s,\infty;\theta^\ast\right)$ for every $s$ (the left side being defined once $r_n^\star>\vert s-\omega_0\vert$, and thus for all large $n$), which is the linear function above. The omitted details are available upon request.
\end{proof}

\begin{proof}[Proof of Corollary \ref{cor:polarization} (in Section \ref{sec:pop})]
	Condition on agent $i$'s prior expectation $\omega_0^i$. Since the agents' beliefs are calibrated, $\mu_0^i$ is the objective distribution of $\omega$ given $\omega_0^i$, and the signal technology $T_rf\left(s\vert\omega\right)$ is common and depends on the state alone. As such, the conditional law of $\left(\omega,s_i\right)$---and thus expected squared loss---is the same under agent $i$'s beliefs as under the objective measure. Given a signal drawn from radius $r$, agent $i$'s optimal action is her posterior expectation $\overline{\mu_{1\vert r}^i}$ (Eq.	\ref{eq:optimala}), and her optimal radius $r_i^\star$ is chosen from a set that includes $r=\infty$. So $\E\left[\left(\overline{\mu_{1\vert r_i^\star}^i}-\omega\right)^2 \vert\,\omega_0^i\right]\le \E\left[\left(\overline{\mu_{1\vert \infty}^i}-\omega\right)^2 \vert\,\omega_0^i\right]$, with strict inequality whenever $\omega_0^i$ is a
	prior expectation for which censoring yields strictly higher expected utility than not censoring. Taking expectations over $\omega_0^i$ and averaging over agents yields $\E\left[\frac{1}{N}\sum_i\left(\overline{\mu_{1\vert r_i^\star}^i}
	-\omega\right)^2\right]\le
	\E\left[\frac{1}{N}\sum_i\left(\overline{\mu_{1\vert \infty}^i}
	-\omega\right)^2\right]$. If the set of such prior expectations has positive probability, the conditional inequality is strict on a set of positive measure, so the unconditional inequality is strict.
\end{proof}

\begin{proof}[Proof of Corollary \ref{cor:ineq} (in Section \ref{sec:pop})]
			Fix the agent's prior expectation $\omega_0=0$, without loss, and fix a finite radius $\bar{r}$. Let $\bar{a}\equiv\sup_{\vert s\vert\leq\bar{r}}\vert a^\star\left(s,\bar{r}\right)\vert$ be the largest magnitude of any action the agent might take under censoring to $\bar{r}$.
		
		We first verify that $\bar{a}<\infty$. The normalizer in Equation \ref{eq:truncatedSignal} does not depend on $s$, so Equation \ref{eq:posterior} factors as $\mu_1\left(\omega\vert s,\bar{r}\right)\propto\left[\mu_0\left(\omega\right)/P_{\bar{r}}\left(\omega\right)\right]f\left(s\vert\omega\right)$, where $P_{\bar{r}}\left(\omega\right)\equiv F\left(\bar{r}\vert\omega\right)-F\left(-\bar{r}\vert\omega\right)$: censoring amounts to updating the tilted prior $\mu_0/P_{\bar{r}}$ with the untruncated likelihood. The signal thus enters only through $f\left(s\vert\omega\right)$, which is strictly positive, bounded, and continuous. As such, the posterior varies continuously in $s$ across the window, with no degeneracy as $s$ approaches its boundary. By the integrability of $\left(1+\vert\omega\vert\right)\mu_0\left(\omega\right)\sup_{\vert s\vert\leq\bar{r}}T_{\bar{r}}f\left(s\vert\omega\right)$ (footnote \ref{fn:integrability}), dominated convergence makes both integrals defining $a^\star\left(\cdot,\bar{r}\right)$ in Equation \ref{eq:optimala} continuous in $s$, with strictly positive denominator. As a continuous function on a compact interval, $a^\star\left(\cdot,\bar{r}\right)$ is bounded, so $\bar{a}<\infty$.
		
		Now consider a true state $\omega>\bar{a}$ (the case $\omega<-\bar{a}$ is symmetric). Whatever signal the censored agent draws, her action lies in $\left[-\bar{a},\bar{a}\right]$, so her loss is at least $\left(\omega-\bar{a}\right)^2$, and thus $V\left(\omega,\bar{r}\right)\geq\left(\omega-\bar{a}\right)^2$. The expected benefit of not censoring, conditional on $\omega$, is then at least
		\begin{equation*}
			V\left(\omega,\bar{r}\right)-V\left(\omega,\infty\right) \geq \left(\omega-\bar{a}\right)^2-V\left(\omega,\infty\right) = \omega^2\left[\left(1-\frac{\bar{a}}{\omega}\right)^2-\frac{V\left(\omega,\infty\right)}{\omega^2}\right].
		\end{equation*}
		As $\omega\rightarrow\infty$, the first term in brackets converges to one while, by Condition \ref{eq:ineqcond}, the second is eventually bounded above by a constant strictly below one. The bracketed expression is therefore eventually bounded below by a strictly positive constant, so the right-hand side diverges. Thus conditional on a true state far enough from the agent's prior expectation, uncensored sampling provides strictly higher expected utility than censoring to any finite radius $\bar{r}$, including $r^\star$.
\end{proof}

\printbibliography

\end{document}